\documentstyle[aps,graphicx]{revtex}
\begin{document}
\draft

\title{Monte Carlo Studies of the Ordering of the Three-Dimensional\\
     Isotropic Heisenberg Spin Glass in Magnetic Fields}
\author{Daisuke Imagawa\thanks{E-mail:imag@spin.ess.sci.osaka-u.ac.jp}
and Hikaru Kawamura\thanks{E-mail:kawamura@ess.sci.osaka-u.ac.jp}}
\address{Department of Earth and Space Science, Faculty of Science,
     Osaka University, Toyonaka, Osaka 560-0043, Japan}
\date{\today}
%
\maketitle

\def\vecS{\mbox{\boldmath$S$}}
\def\unitvec{\hat{\mbox{\boldmath$e$}}}

%
%
%
\begin{abstract}

Spin and chirality orderings of the three-dimensional Heisenberg
spin glass under magnetic fields are studied by large-scale equilibrium
Monte Carlo simulations.
It is found
that the chiral-glass transition
and the chiral-glass ordered state,
which are essentially of the same character
as their zero-field counterparts,
occur under magnetic fields.
The chiral-glass ordered state exhibits a one-step-like
peculiar replica-symmetry breaking in the chiral sector, while it
does not accompany the spin-glass order perpendicular to the applied field.
Critical perperties of the chiral-glass transition are different
from those of the standard Ising spin glass.
Magnetic phase diagram of the model is constructed, which reveals that
the chiral-glass state is quite robust against magnetic fields.
The chiral-glass transition line has  a character of the
Gabay-Toulouse line of the mean-field model, yet its physical origin
being entirely different.
These numerical results are discussed in light of the recently developed
spin-chirality decoupling-recoupling scenario.
Implication to experimental phase diagram is also discussed.

\end{abstract}


%

%
%
%
%
%

\section{Introduction}

   In the studies of spin glasses, much effort has been devoted
either exprimentally or theoretically
to the properties  under magnetic fields.
Unfortunately, our understanding of them
still has remained unsatisfactory\cite{SGrev}.
On theoretical side, most of the numerical studies
have  focused on the properties of the simple Ising model,
especially the three-dimensional (3D) Edwards-Anderson (EA) model.
While the existence of a true thermodynamic spin-glass (SG) transition
has been established  for this model in zero field,
the question of its existence or nonexistence in magnetic fields
has remained unsettled.
This question is
closely related to the hotly debated issue of whether
the ordered state of the 3D Ising SG  in zero field
exhibits a replica-symmetry breaking (RSB) or not.

If one tries to understand real
experimental SG ordering,
one has to remember that many of real SG materials are
more or less Heisenberg-like rather than Ising, in the sense that
the random magnetic anisotropy is considerably weaker than
the isotropic exchange interaction\cite{SGrev,OYS}.
For example,
in widely studied canonical spin glasses, {\it i.e.\/},
dilute metallic alloys such as AuFe, AgMn and CuMn,
random magnetic anisotropy originated from the
Dzyaloshinski-Moriya interaction or the dipolar interaction is
often one or two magnitudes weaker than the isotropic RKKY
interaction.
Numerical simulations
have indicated that the
isotropic 3D Heisenberg SG with finite-range interaction
does not exhibit the conventional
SG order at finite
temperature in zero field
\cite{SGrev,OYS,Matsubara1,Kawamura92,Yoshino,Kawamura95,Kawamura98,HK1}.
(However, see also Ref.\cite{Matsubara2}.)
Since applied fields
generally tend to suppress the SG ordering,
a true thermodynamic SG transition is even more unlikely
under magnetic fields in case of the 3D Heisenberg SG.

Experimentally, however, a rather sharp transition-like behavior
has been observed under magnetic fields
in typical Heisenberg-like SG magnets,
although it is not completely clear whether
the observed anomaly corresponds
to a true thermodynamic transition\cite{SGrev,Orbach,Campbell}.
The situation is in contrast to  the zero-field case
where the existence of a true thermodynamic SG transition has
been established experimentally\cite{SGrev}.
Set aside the question of 
the strict nature of the SG ``transition'', it is experimentally
observed that  a weak applied field
suppresses the zero-field SG transition temperature rather quickly.
For higher fields, the SG ``transition''
becomes much more robust to fields,
where the ``transition temperature'' shows much less field dependence
\cite{SGrev,Orbach,Campbell}.
Such  behaviors of the SG transition temperature under magnetic fields
$T_g(H)$ were often interpreted in terms of the mean-field model
\cite{SGrev,Orbach}. Indeed,
the mean-field Sherrington-Kirkpatrick (SK) model\cite{SK}
with
an infinite-range Heisenberg exchange interaction with weak
random magnetic anisotropy
exhibits a transition line
similar to the experimental one \cite{Kotliar},
{\it i.e.\/}, the so-called de Almeida-Thouless (AT) line\cite{AT}
$H\propto (T_g(0)-T_g(H))^{3/2}$
in weak-field regime where the anisotropy is important,
and the Gabay-Toulouse (GT) line\cite{GT}
$H\propto (T_g(0)-T_g(H))^{1/2}$ in strong-field regime
where the anisotropy is unimportant.
Nevertheless, if one notes that the true finite-temperature transition under
magnetic fields, though
possible in the infinite-range SK model, 
is unlikely to occur
in a more realistic finite-range Heisenberg model,
an apparent success of the mean-field model in explaining the
experimental phase diagram should be taken with strong reservation.
Thus, the question of the true nature of the experimentally observed
SG ``transition'' under magnetic fields remains unsolved.

\begin{figure}
\begin{center}
\includegraphics{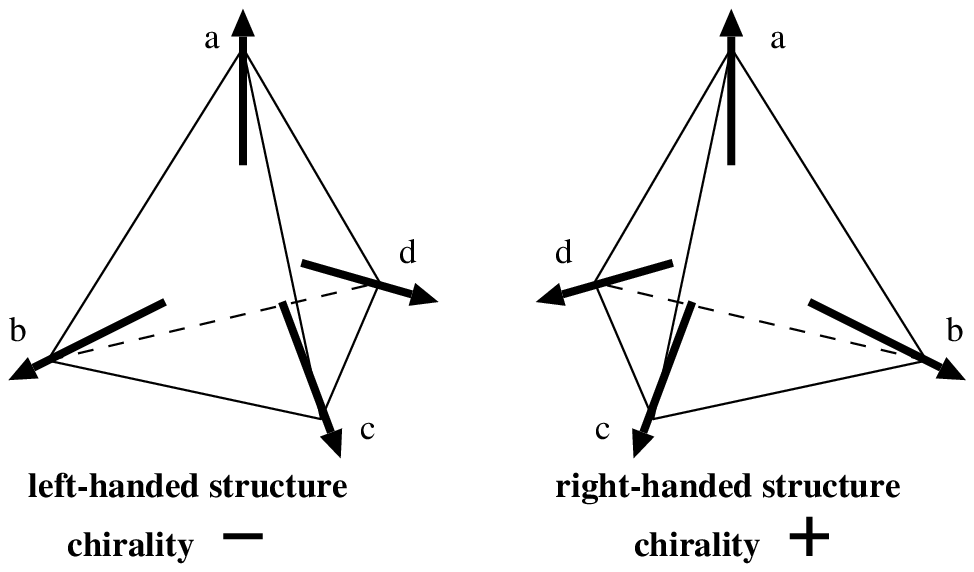}
\caption{Two energetically degenerate
``chiral'' structures characterized by the mutually
opposite sign of the chirality. The labels $a$-$d$ denote four
distinct Heisenberg spins.
}
\label{fig-chiral}
\end{center}
\end{figure}

Recently, one of the present authors has proposed a scenario,
the spin-chirality decoupling-recoupling scenario, aimed at explaining
some of the puzzles
concerning the experimentally observed SG transition\cite{Kawamura92}.
In this scenario, {\it chirality\/}, which is a multispin variable
representing  the sense or the handedness of local
noncoplanar spin structures induced by spin frustration,
plays an essential role.  As  illustrated in
Fig.\ref{fig-chiral}, locally
noncoplanar spin structures inherent to the
SG ordered state sustain two energetically degenerate ``chiral'' states,
``right-handed'' and ``left-handed'' states, characterized by mutually
opposite signs of the ``chiralities''.  Here, one may
define the local chirality
by {\it three\/} neighboring Heisenberg spins 
$\vecS_1$, $\vecS_2$ and $\vecS_3$ by,
\begin{equation}
\chi = \vecS_1\cdot(\vecS_2\times\vecS_3)\ \ .
\label{chidef}
\end{equation}
This type of chirality is called ``scalar chirality'',
in distinction with ``vector chirality'' defined as a vector product of
two neighboring Heisenberg spins,
$\vecS_1\times\vecS_2$\cite{Kawamiya}.
Note that the chirality defined by Eq.(\ref{chidef})
is a pseudoscalar in the sense that
it is invariant under global $SO(3)$
spin rotations but changes its sign  under $Z_2$ spin reflections
(or inversions which can be viewed as a
combination of reflections and rotations).

For a fully isotropic Heisenberg SG, in particular,
the chirality scenario of Ref.\cite{Kawamura92}
claims the occurrence of a novel {\it chiral-glass\/}
ordered state in which only the chirality exhibits a glassy long-range
order (LRO) while the spin remains paramagnetic.
At the chiral-glass transition,
among the global symmetries of the Hamiltonian,
$O(3)=Z_2\times SO(3)$,
only the $Z_2$ spin reflection (inversion)
symmetry is broken spontaneously
with keeping the $SO(3)$ spin rotation symmetry preserved.
Note that this picture
entails the spin-chirality (or $SO(3)-Z_2$)
decoupling on long length and time scales:
Namely, although the chirality is not independent of the spin
on microscopic length scale,
it eventually exhibits a long-distance behavior
entirely different from the spin.
Such a chiral-glass transition
without the conventional spin-glass order
was indeed observed
in  recent equilibrium  and off-equilibrium
Monte Carlo (MC) simulations  in zero field performed by Hukushima and one of
the authors (H.K.)\cite{Kawamura98,HK1}.
It was also found there
that the critical properties associated with the chiral-glass transition
were different from those of the Ising SG, and that the chiral-glass
ordered state exhibited a one-step-like peculiar RSB.

In the chirality scenario of Ref.\cite{Kawamura92},
experimental SG
transition in real Heisneberg-like SG magnets is regarded essentially as
a chiral-glass transition 
``revealed'' via the random magnetic anisotropy.
Weak but finite random magnetic anisotropy inherent to real 
magnets
``recouples'' the spin to the chirality,
and the chiral-glass transition shows up
as an experimentally observable
{\it spin\/}-glass transition
in real Heisneberg-like SG magnets.
An interesting outcome of this picture is that
the experimental SG transition is dictated 
by the chiral-glass transition of the fully isotropic system, {\it not
by the spin-glass transition of the fully isotropic system\/}, which
has been separated from the chiral one.

Very recently, the present authors discussed some of the
possible
consequences of the chirality scenario of Ref.\cite{Kawamura92}
on the finite-field properties of
the fully isotropic 3D Heisenberg SG\cite{KawaIma}.
It was argued there that
the chiral-glass transition, essentially
of the same character as the  zero-field one,
occurred also in finite fields.
In the weak field regime, the transition line was
predicted to behave as
\begin{equation}
T_{\rm CG}(0)-T_{\rm CG}(H)=cH^2 + dH^4 + \cdots\ \ ,
\label{eqn:phaseline}
\end{equation}
where $c$ and $d$ are constants.
Generally, the coefficient $c$ could be either
positive or negative. An interesting observation here is that
the chiral-glass transition line (\ref{eqn:phaseline})
aparrently has a form similar to
the GT line of the mean-field model.
We emphasize, however, that
their physical origin is entirely different.
The quadratic dependence of the chiral-glass transition line
is simply of regular origin, whereas that of the GT-line in the
SK model cannot be regarded so.

In the present paper, we report on our results of
large-scale Monte Carlo simulations on the 3D isotropic Heisenberg SG, 
performed with the aim to
reexamine the SG ordering in magnetic fields
in light of the chirality scenario. In particular, by
means of extensive numerical
simulations, we wish to
clarify in detail
how the spin and the chirality order in applied fields.
Part of the MC results have been reported in Ref.\cite{KawaIma}.


The present paper is organized as follows.
In \S\ref{secModel}, we introduce our model and explain some of the
details of our numerical method. Various physical
quantities calculated in our MC simulations are defined in \S\ref{secPhysQ}.
The results of MC simulations are presented in \S\ref{secResult}.
The results for
the chirality- and spin-related quantities are presented
in \S\ref{subsecChiral} and \S\ref{subsecSpin},
respectively.
It is found that the chiral-glass
transition, essentially of the same character as the zero-field one,
occurs under magnetic fields.
The chiral-glass ordered state exhibits a one-step-like
peculiar replica-symmetry breaking in the chiral sector, while it
does not accompany the spin-glass order perpendicular to the applied field.
Critical properties of the chiral-glass
transition are analyzed in \S\ref{subsecCritical}.
The analysis suggests
that the universality class
of both the zero-field and finite-field chiral-glass transitions
might be common, which, however,
differs from that of the
standard 3D Ising SG. In \S\ref{subsecPhase},
we construct a
magnetic phase diagram of the model.
The chiral-glass ordered state
remains quite robust against magnetic fields, while
the chiral-glass transition line in applied fields has a
character of the GT line of the mean-field model.
Section \ref{summary} is devoted to summary and
discussion. Our numerical results are
discussed in terms of the recent experimental result on
canonical SG.

%
%
%
\newpage

\section{The model and the method}
\label{secModel}

In this section, we introduce our model and explain some of
the details of our numerical method.
The model we consider is the isotropic classical Heisenberg
model on a 3D simple cubic lattice defined by the
Hamiltonian,
\begin{equation}
{\cal H}=-\sum_{<ij>}J_{ij}\vecS_i\cdot\vecS_j - H\sum_iS_i^z,
\label{hamil}
\end{equation}
%
where $\vecS_i=(S_i^x,S_i^y,S_i^z)$ is a three-component unit vector,
and $H$ is the intensity of magnetic field applied along
the $z$ direction.
The  nearest-neighbor coupling $J_{ij}$ is assumed to
take either
the value $J$ or $-J$ with equal probability ($\pm J$ distribution).

We perform equilibrium MC simulations on this model.
Simulations are performed for a variety of fields
$H/J=0.05$, 0.1, 0.5, 2.0, 3.0, and 5.0, while most extensive calculations
are performed for $H/J=0.1$ and 0.5.
The lattices studied are simple-cubic lattices with $N=L^3$ sites
with $L=6$, 8, 10, 12 and 16 with periodic boundary conditions.
Sample average is taken over 128-1400 independent bond realizations,
depending on the system size $L$ and the field intensity $H$.
Limited amount of data is also taken for $L=20$ in some cases
  (30 samples only) to check
the size dependence of some physical quantities.
To facilitate efficient thermalization, we combine the standard
heat-bath method  with the temperature-exchange technique\cite{HN}.
Care is taken to be sure that
the system is fully equilibrated.
Equilibration is checked by the following procedures:
First, we monitor the system to travel back and forth
many times during the
the temperature-exchange process (typically more than 10 times)
between the maximum and minimum temperature points, and at the same time check
that the relaxation
due to the standard heat-bath updating
is reasonably fast at the highest temperature,
whose relaxation time is of order $10^2$ Monte Carlo steps
per spin (MCS). This guarantees that  different parts of
the phase space are sampled in each ``cycle'' of the temperature-exchange
run. Second, we check
the stability of the results against at least three times longer runs
for a subset of samples. Error bars of
physical quantities are estimated by the sample-to-sample statistical
fluctuation over bond realizations.
Further details of our Monte Carlo simulations
are given in Table \ref{table-condition}.
%
%
%
%
%
%
\newpage
\begin{table}
\begin{center}
\caption{Details of our MC simulations.
$H/J$ represents the magnetic-field intensity, $L$ the lattice size,
$N_{\rm s}$ the total number of samples,  $N_{\rm T}$ the total number of
temperature points used in the temperature-exchange run,
$T_{\max}/J$ and $T_{\min}/J$  the maximum and minimum
temperatures in the temperature-exchange run.}
\label{table-condition}
\scalebox{0.5}{\includegraphics[width=\columnwidth]{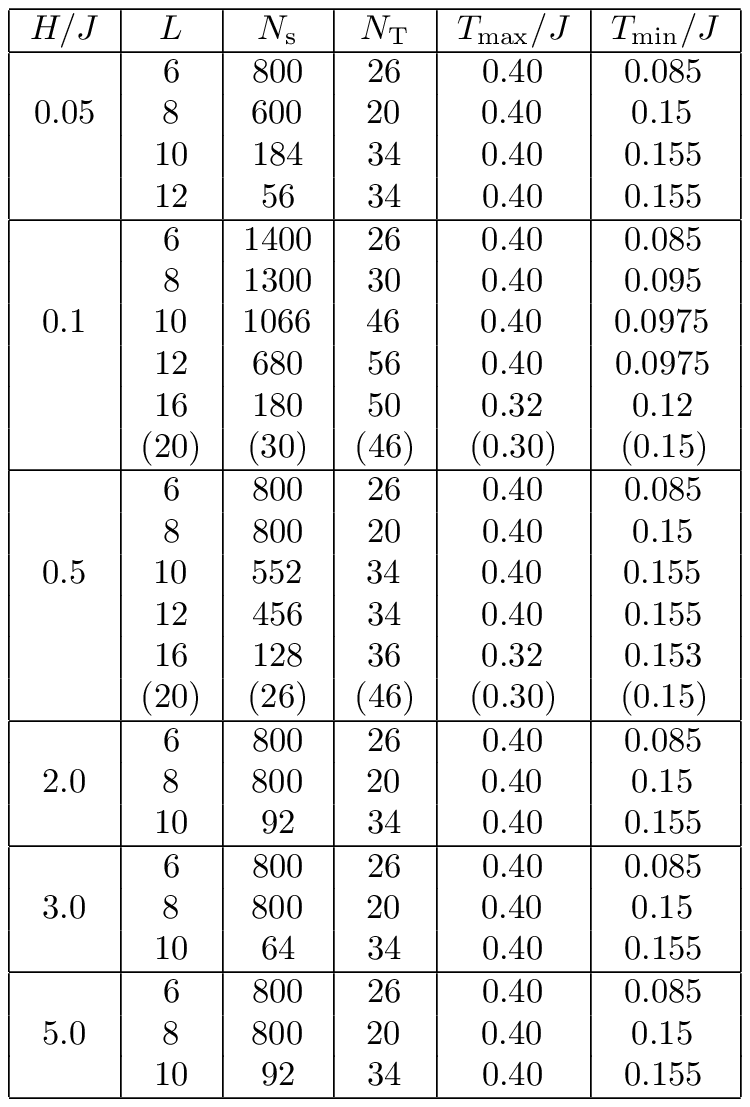}}
\end{center}
\end{table}

%
%
\newpage
\section{Physical Quantities}
\label{secPhysQ}

In this section, we
define various physical quantities calculated in our
simulations below.
\label{defquantity}
\subsection{Chirality-related quantities}

Let us begin with the definition of the chirality.
We define
the local chirality at the $i$-th site and in the $\mu$-th
direction, $\chi_{i\mu}$, for three neighboring Heisenberg spins
by the scalar
\begin{equation}
\chi_{i\mu}=
\vecS_{i+\unitvec_{\mu}}\cdot
(\vecS_i\times\vecS_{i-\unitvec_{\mu}}),
\end{equation}
%
where $\unitvec_{\mu}\ (\mu=x,y,z)$ denotes a unit vector along the
$\mu$-th axis. By this definition, there are in total $3N$ local
chiral variables in the system.

The mean local amplitude of the chirality, $\bar \chi$, may be defined by
\begin{equation}
\bar{\chi}^2=\frac{1}{3N}\sum_{i=1}^N
\sum_{\mu=x,y,z}[\langle\chi_{i\mu}^2\rangle]\ \ ,
\end{equation}
where  $\langle\cdots\rangle$ represents the thermal average and
[$\cdots$] the average over the bond disorder.
This quantity vanishes for coplanar spin structures, and its magnitude
tells us the extent of the noncoplanarity of the local spin
structures.

By considering two independent systems (``replicas'') described by
the same Hamiltonian (\ref{hamil}),
one can define an overlap of the chiral variable
via the relation,
\begin{equation}
q_{\chi}=
\frac{1}{3N}\sum_{i=1}^N\sum_{\mu=x,y,z}\chi_{i\mu}^{(1)}\chi_{i\mu}^{(2)}\ \ ,
\end{equation}
where $\chi_{i\mu}^{(1)}$ and  $\chi_{i\mu}^{(2)}$ represent the chiral
variables of the replicas 1 and 2, respectively.
In our simulations, we prepare the two replicas 1 and 2 by
running two independent sequences of  systems
in parallel with different spin initial conditions and
different sequences of random numbers.
In terms of this chiral overlap $q_{\chi}$, the chiral-glass
order parameter may be defined by
\begin{equation}
q_{\chi}^{(2)}=[\langle q_{\chi}^2\rangle]\ \ ,
\end{equation}
while the associated chiral-glass susceptibility may be defined by
\begin{equation}
\chi_\chi=3N[\langle q_{\chi}^2\rangle]\ \ .
\end{equation}
Unlike the spin variable, the local magnitude
of the chirality is temperature dependent somewhat.
In order to take account of this short-range order effect,
we also consider the reduced chiral-glass order parameter
$\tilde q_\chi^{(2)}$
and the reduced chiral-glass susceptibility $\tilde \chi_\chi$
by dividing $q_\chi^{(2)}$ and $\chi_\chi$
by appropriate powers of $\bar \chi$,
\begin{equation}
\tilde q_\chi^{(2)}=\frac{q_\chi^{(2)}}{\bar \chi^4},\ \ \ \
\tilde \chi_\chi=\frac{\chi_\chi}{\bar \chi^4}.
\end{equation}
The Binder ratio of the chirality is defined by
\begin{equation}
g_{\chi}=
\frac{1}{2}
\left(3-\frac{[\langle q_{\chi}^4\rangle]}
{[\langle q_{\chi}^2\rangle]^2}\right)\ \ .
\end{equation}
One may also
define  the distribution function of the chiral overlap
$q_{\chi}$ by
\begin{equation}
P_\chi(q^{\prime}_{\chi})=[\langle\delta(q_\chi^{\prime}-q_{\chi})\rangle]\ \ .
\end{equation}

  In order to study the equilibrium dynamics of the model,
we also compute the autocorrelation function of the chirality defined by
\begin{equation}
C_{\chi}(t)=\frac{1}{3N}\sum_{i=1}^N\sum_{\mu=x,y,z}
[\langle\chi_{i\mu}(t_0)\chi_{i\mu}(t+t_0)\rangle ]\ \ .
\label{Cxt}
\end{equation}
where the ``time'' $t$ is measured in units of MCS.
In computing (\ref{Cxt}), simulation is performed
according to the standard heat-bath updating without the
temperature-exchange procedure, while
the starting spin configuration at $t=t_0$ is taken from
the equilibrium spin configurations
generated in our temperature-exchange MC runs.

We also calculate the so-called $G$ and $A$ parameters
for the chirality, recently discussed in the
literature\cite{Marinari98,Bokil,Marinari99,Parisi99,HK2,Picco,Ritort},
defined by,
\begin{equation}
G_{\chi}=
\frac{[\langle q_{\chi}^2\rangle ^2]-[\langle q_{\chi}^4\rangle]}
{[\langle q_{\chi}^2\rangle]^2-[\langle q_{\chi}^4\rangle]},
\end{equation}
\begin{equation}
A_{\chi}=
\frac{[\langle q_{\chi}^2\rangle ^2]-[\langle q_{\chi}^4\rangle]}
{[\langle q_{\chi}^2\rangle]^2}\ \ .
\end{equation}
These $G_\chi$ and $A_\chi$
parameters are closely related to the sample-to-sample
fluctuation of the chiral order parameter. The
$A$ parameter is known to be an indicator of the non-self-averagingness of
the order parameter, {\it i.e.\/},
it vanishes in the state where the order parameter
is self-averaging and
takes a nonzero value otherwise\cite{Marinari99}.
By contrast, the $G$ parameter could take a nonzero value even in a
self-averaging ordered state, and hence, cannot be used as an unambiguous
indicator of the non-self-averagingness\cite{Bokil}.
However, since in the thermodynamic limit
it vanishes in the high-temperature phase
and takes a nonzero value in the ordered
state, it can still be used as an indicator of a phase transition.

\subsection{Spin-related quantities}

As in the case of the chirality, it is convenient to define an overlap
variable  for the
Heisenberg spin. In this case, the overlap
might naturally be defined
as a {\it tensor\/} variable $q_{\mu\nu}$
between the $\mu$ and $\nu$
components  ($\mu$, $\nu$=$x,y,z$) of the Heisenberg spin,
\begin{equation}
q_{\mu\nu}=\frac{1}{N}\sum_{i=1}^N  S_{i\mu}^{(1)}S_{i\nu}^{(2)}\ \ ,
\ \ (\mu=x,y,z),
\end{equation}
%
where $\vecS_i^{(1)}$ and $\vecS_i^{(2)}$ are the $i$-th
Heisneberg spins of the replicas 1 and 2, respectively.
In terms of these tensor overlaps, the ``longitudinal'' (parallel to the
applied field) and ``transverse'' (perpendicular to the applied field)
SG order parameters may be defined by
\begin{equation}
q_{\rm L}^{(2)} = [\langle q_{\rm L}^2\rangle],
\ \ \ \ q_{\rm L}^2 = q_{zz}^2,
\end{equation}
\begin{equation}
q_{\rm T}^{(2)} = [\langle q_{\rm T}^2\rangle],\ \ \ \
q_{\rm T}^2 = \sum_{\mu,\nu = x, y}q_{\mu\nu}^2\ \ .
\end{equation}
The associated  longitudinal and transverse Binder
ratios are  defined by
\begin{equation}
g_{\rm L} = \frac{1}{2}
\left(3 - \frac{[\langle q_{\rm L}^4\rangle]}
{[\langle q_{\rm L}^2\rangle]^2}\right),
\label{binL}
\end{equation}
\begin{equation}
g_{\rm T} = \frac{1}{2}
\left(6 - 4\frac{[\langle q_{\rm T}^4\rangle]}
{[\langle q_{\rm T}^2\rangle]^2}\right)\ \ .
\end{equation}
Here, $g_{\rm L}$ and $g_{\rm T}$
are normalized so that, in the thermodynamic limit,
they vanish in the high-temperature phase and gives unity in the
nondegenrate ordered state.

Since an odd quantity like $\langle q_{\rm L}\rangle$
does not vanish in applied fields, one can also define
the ``connected Binder ratio'' for the
longitudinal component\cite{Picco},
\begin{equation}
g^{\prime}_{\rm L}=\frac{1}{2}\left(
3-\frac{[\langle(q_{\rm L}-\langle q_{\rm L}\rangle)^4\rangle]}
{[\langle(q_{\rm L}-\langle q_{\rm L}\rangle)^2\rangle]^2}\right).
\end{equation}
In applied fields, $g^{\prime}_{\rm L}$ might well behave differently from
$g_{\rm L}$.

The spin-overlap distribution function is generally defined in the
tensor space. In the following, we pay  particular attention
to its transverse ($XY$) part. The relevant transverse overlap
originally has $2^2=4$ independent components.
For the convenience of illustration,
we follow Ref.\cite{KawaLi} here and introduce
the projected transverse-spin-overlap
distribution function $P_{\rm T}(q_{{\rm diag}})$
defined in terms of the diagonal
overlap $q_{{\rm diag}}$ which is the trace of the tensor overlap
$q_{\mu\nu}$'s,
\begin{equation}
q_{{\rm diag}}=\sum _{\mu=x,y} q_{\mu \mu}
          =\frac{1}{N}\sum_{i=1}^N (S_{ix}^{(1)}S_{ix}^{(2)}
           +S_{iy}^{(1)}S_{iy}^{(2)}).
\end{equation}
The distribution function $P_{\rm T}(q_{{\rm diag}})$ is symmetric
with respect to $q_{{\rm diag}}=0$.
In the high-temperature phase,
each $q_{\mu\nu}$ ($\mu, \nu=x,y$) is expected
to be Gaussian-distributed around
$q_{\mu\nu}=0$ in the $L\rightarrow \infty$ limit, and so is
$q_{{\rm diag}}$.

\begin{figure}
\begin{center}
\scalebox{1.3}{\includegraphics{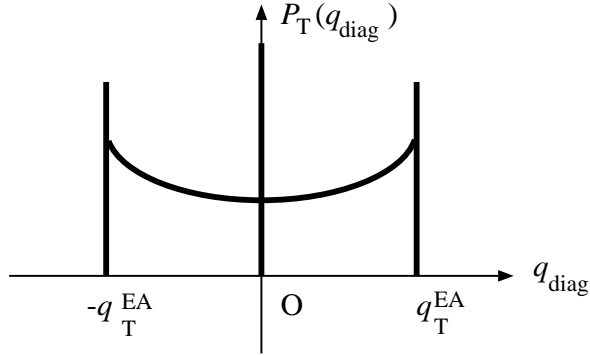}}
\caption{Sketch of the form of the
transverse diagonal-spin-overlap distribution function
$P_{\rm T}(q_{{\rm diag}})$
in the thermodynamic limit,
expected when there
exists a finite transverse SG long-range order with a nonzero
$q_{\rm T}^{{\rm EA}}>0$.
}
\label{XYLRO}
\end{center}
\end{figure}

Let us hypothesize here that there exists a transverse
{\it spin\/}-glass ordered state
characterized by a
nonzero $q_{\rm T}^{(2)}$, or by a nonzero
EA transverse SG order parameter
$q_{\rm T}^{{\rm EA}}>0$.
Reflecting the fact that $q_{{\rm diag}}$ transforms nontrivially
under independent $O(2)$ rotations
around the $z$-axis  on the two replicas, which are
the symmetries relevant to the
transverse spin components in the presence of magnetic fields,
even a self-overlap   contributes nontrivial weights to
$P_{\rm T}(q_{{\rm diag}})$
other than at $\pm q_{\rm T}^{{\rm EA}}$.
In the $L\rightarrow \infty$ limit,
the self-overlap part of $P_{\rm T}(q_{{\rm diag}})$ should be given
by
\begin{equation}
P_{\rm T}(q_{{\rm diag}})= \frac{1}{2}\delta (q_{{\rm diag}})
           +\frac{1}{2\pi}
            \frac{1}
            {\sqrt{(q_{\rm T}^{{\rm EA}})^2-q_{{\rm diag}}^2}}.
\label{Ptform}
\end{equation}
The derivation of Eq.(\ref{Ptform}) has been given in Ref.\cite{KawaLi}
in the context of the {\it XY\/} SG.
If the transverse SG ordered state accompanies RSB,
the associated nontrivial contribution would be added to
the one given by Eq.(\ref{Ptform}). In any case, an important observation
here is that, as long as the ordered state possesses a
finite transverse
SG LRO, the diverging peak should arise in $P_{\rm T}(q_{{\rm diag}})$
at $q_{{\rm diag}}=\pm q_{\rm T}^{{\rm EA}}$ as illustrated in Fig.\ref{XYLRO},
irrespective to the occurrence of
the RSB.
%
%
%
%
%
%
%
%
%

\newpage
\section{Monte Carlo Results}
\label{secResult}

This section is the core part of the present paper.
Here,  we  present our MC results on the 3D
$\pm J$ Heisenberg SG in magnetic fields.

\subsection{Chirality-related quantities}
\label{subsecChiral}
\begin{figure}
\begin{center}
\includegraphics[width=\linewidth]{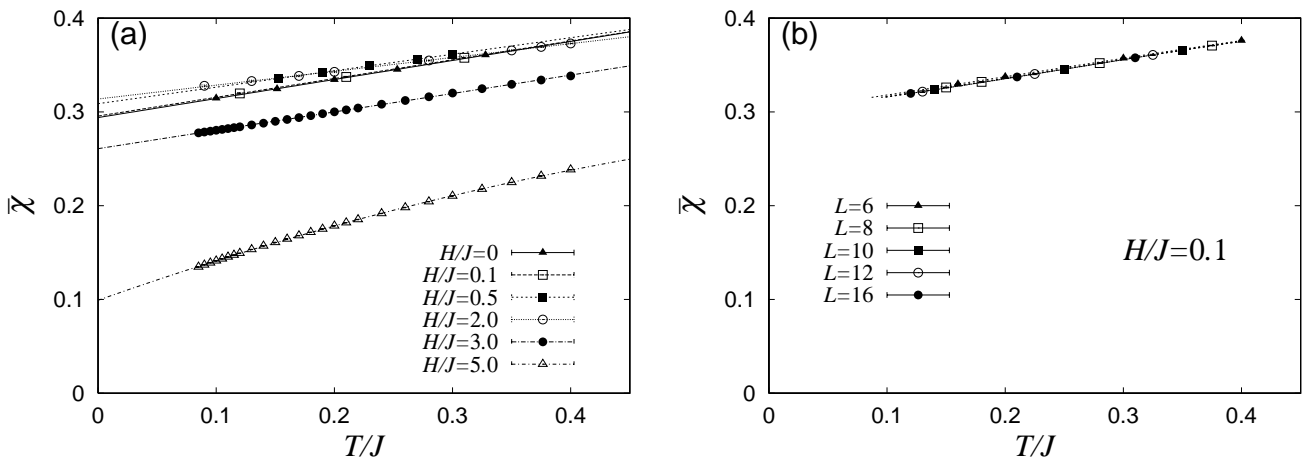}
\caption{Temperature dependence of the mean local amplitude
of the chirality for various magnetic fields. The lattice size is
$L=16$ for $H/J=0$, 0.1 and 0.5, and is $L=6$ for other field values.
For the case of $H/J=0.1$, the size dependence of $\bar \chi$
is shown in Fig.(b).
}
\label{fig-locx}
\end{center}
\end{figure}

First, we begin with the chirality-related quantities.
In Fig.\ref{fig-locx},
we show the temperature and size dependence of the
mean local amplitude of the chirality  for various
fields. As can clearly be seen from Fig.\ref{fig-locx}(a),
extrapolation of $\bar{\chi}(T)$ to $T=0$ gives
non-zero values as long as the applied field intensity is not too
large, {\it i.e.\/}, $\bar{\chi}(T\rightarrow 0)
\simeq 0.294$, 0.295, 0.308, 0.313, 0.260, and 0.100
for $H/J=0$, 0.1, 0.5, 2.0, 3.0, and 5.0, respectively.
This indicates that the spin ordering
of the 3D Heisenberg SG is certainly noncoplanar, which guarantees
that  the system
sustains the nontrivial chirality. Meanwhile, a direct inspection of the
spin pattern suggests that such noncoplanar
spin configurations realized at low temperature in zero and weak fields
is rather close to the coplanar one. Indeed,
for completely random configurations of
Heisenberg spins, $\bar \chi$ should take a value
$\sqrt 2/3\simeq 0.4714\cdots $\cite{Kawamura95}, a value considerably
larger than the extrapolated $\bar\chi(T\rightarrow 0)$ values.
This again suggests that the noncoplanar configuration realized
in zero and weak
fields is close to the coplanar one.
Interestingly, our MC data indicate that, in the weak field regime,
$\bar\chi$
slightly {\it increases\/}
with increasing magnetic field at fixed temperatures.
This observation could be understood if one notes that
the zero-field
noncoplanar spin configuration is close to the
coplanar one, 
and that the
application of a magnetic field  to such nearly coplanar spin configuration
tends to ``rise up'' the spins from this plane with keeping
the plane orthogonal to the
applied field. This gives rise to
more ``three-dimensional'' local spin structures with larger
$\bar \chi$. Of course,
when the field is further increased,
$\bar\chi$ eventually decreases simply because strong enough
fields force spins to align along   the field.
In Fig.\ref{fig-locx}(b),
we show the size dependence of $\bar{\chi}$ for the field $H/J=0.1$. As can
be seen from the figure, there is very little size dependence in $\bar{\chi}$.

\begin{figure}
\begin{center}
\includegraphics[width=\linewidth]{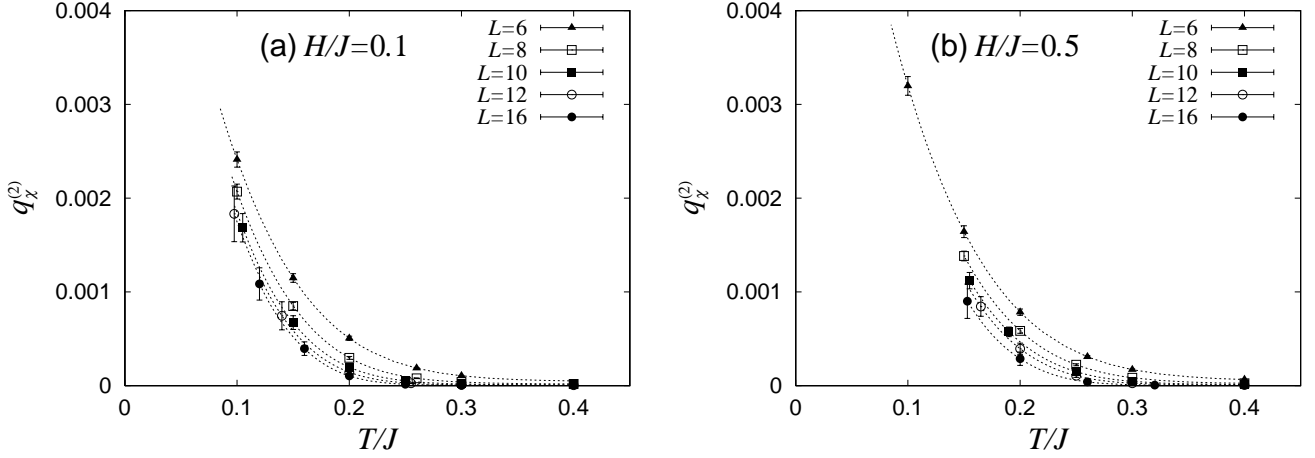}
\caption{Temperature and size dependence of the chiral-glass order parameter
for the field (a) $H/J=0.1$,  and (b) $H/J=0.5$.}
\label{fig-qx2}
\end{center}
\end{figure}

\begin{figure}
\begin{center}
\includegraphics[width=\linewidth]{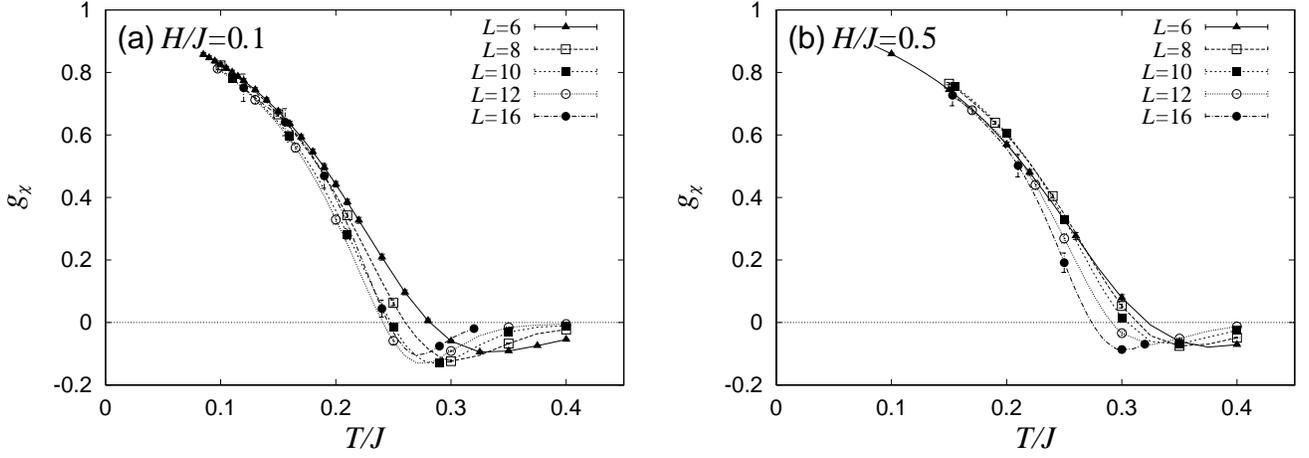}
\caption{Temperature and size dependence of the Binder ratio
     of the chirality for the fields (a) $H/J=0.1$,  and  (b)
$H/J=0.5$.}
\label{fig-gx}
\end{center}
\end{figure}

In Fig.\ref{fig-qx2}, we show the chiral-glass order parameter
$q_{\chi}^{(2)}$ for the fields (a) $H/J=0.1$, and (b) $H/J=0.5$.
For both fields,
$q_{\chi}^{(2)}$ increases rather sharply at lower temperatures.
%
%
In Figs.\ref{fig-gx}(a) and (b), we show the Binder ratio of the chirality
$g_{\chi}$ for the fields (a) $H/J=0.1$,  and (b) $H/J=0.5$.
As can be seen from the figures, $g_\chi $ 
exhibits a negative dip which, with increasing $L$,
tends to deepen and shift toward lower
temperature.  Furthermore, $g_\chi $
of various $L$ cross at a temperature slightly above the
dip temperature
$T_{{\rm dip}}$  {\it on negative side of
$g_\chi$\/},  eventually merging at temperatures lower than $T_{{\rm dip}}$.
The observed  behavior of $g_\chi $ is similar to the one
observed in zero field\cite{HK1}.
As argued in
Ref.\cite{HK1}, the persistence of a negative dip and the crossing occurring
at $g_\chi<0$ are strongly
suggestive of the occurrence of a finite-temperature transition
where $g_\chi (T_{{\rm CG}})$
takes a {\it negative\/} value in the $L\rightarrow \infty $ limit.

\begin{figure}
\begin{center}
\scalebox{0.7}{\includegraphics{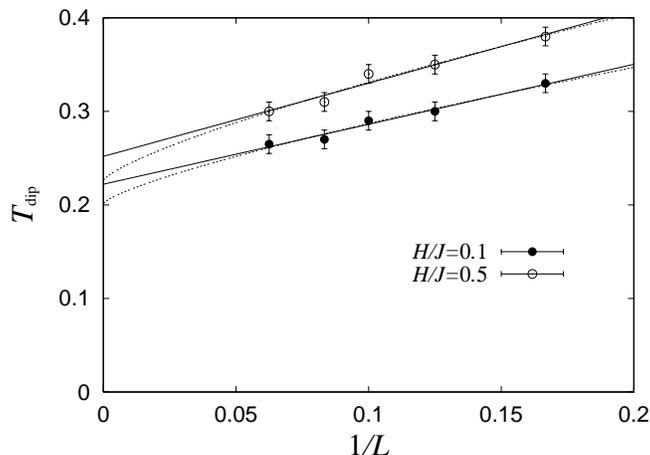}}
\caption{The dip temperature of  $g_\chi$ 
plotted versus $1/L$ for the fields $H/J=0.1$ and 0.5.
$L=\infty$ extrapolation is performed, either based on the  $1/L$ fit
(solid line), or on the $1/L^{1/1.3}$ fit (dashed curve).
The extrapolated values,  $T_{{\rm dip}}(\infty)$, give estimates of the
bulk chiral-glass transition temperature $T_{{\rm CG}}$.}
\label{fig-dip}
\end{center}
\end{figure}

In Fig.\ref{fig-dip}, we plot the  negative-dip
temperature $T_{{\rm dip}}(L)$
versus $1/L$ for the fields $H/J=0.1$ and $H/J=0.5$. For both
fields, the data lie on a
straight line fairly well. The linear extrapolation to $1/L=0$,
as shown by the solid lines in the figure,
gives our first estimates of the bulk chiral-glass
transition temperature, {\it i.e.\/}, $T_{{\rm CG}}/J\simeq 0.23$
for $H/J=0.1$
and $T_{{\rm CG}}/J\simeq 0.25$ for $H/J=0.5$.
More precisely, $T_{{\rm dip}}(L)$ should scale
with $L^{-1/\nu }$ where $\nu $ is the chiral-glass correlation-length
exponent. As shown below, our estimate of $\nu \simeq 1.3$
comes close to unity, more or less
justifying the linear extrapolation employed here.
Indeed, extrapolation with respect
to $L^{-1/1.3}$, shown by the dashed curve in Fig.\ref{fig-dip},
yields $T_{{\rm CG}}/J\simeq 0.21$ for $H/J=0.1$ and
$T_{{\rm CG}}/J\simeq 0.23$ for $H/J=0.5$.

\begin{figure}
\begin{center}
\includegraphics{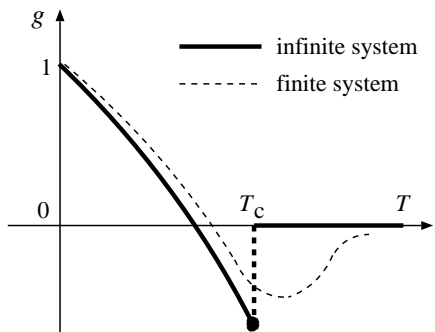}
\caption{Sketch of the behavior of the Binder ratio expected in a system
exhibiting a one-step RSB transition. The solid curve represents the behavior
of an infinite system, while the dashed curve represents that of a finite
system.}
\label{fig-binx}
\end{center}
\end{figure}

As shall be argued below, we attribute several unusual features of $g_\chi$,
{\it e.g.\/}, the growing negative dip  and the crossing
occurring at $g_\chi<0$,
to the possible one-step-like peculiar RSB in the chiral-glass ordered state.
In  systems exhibiting the one-step RSB, {\it e.g.\/},
the mean-field three-state Potts glass,
the Binder ratio
is known to behave
as illustrated in Fig.\ref{fig-binx}, with a
negative dip and the crossing occurring on the negative side\cite{HK2}.
Indeed, such
a behavior
is not
dissimilar to the one we have observed in Fig.\ref{fig-gx}.
%
%
%

\begin{figure}
\begin{center}
\includegraphics[width=\linewidth]{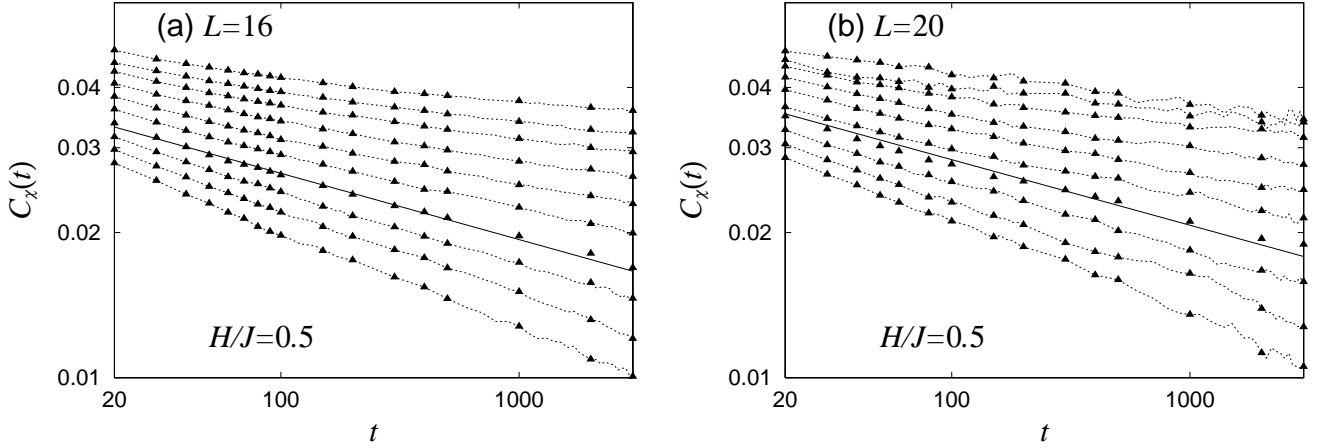}
\caption{Temporal decay of the equilibrium
chiral autocorrelation function at various temperatures
for the field $H/J=0.5$ for the sizes (a) $L=16$, and (b) $L=20$.
Temperatures correspond to $T/J=0.17$, 0.18, 0.19, 0.20,
0.21, 0.22, 0.23, 0.24, 0.25, 0.26 from top to bottom.
Straight lines of power-decay fit are shown in both figures (a) and (b) 
at $T/J=0.23$.
}
\label{fig-auto}
\end{center}
\end{figure}

An independent estimate of
$T_{{\rm CG}}$ can  be obtained from
the equilibrium dynamics of the model.
Thus, we also calculate  the chirality autocorrelation
function  $C_\chi (t)$ defined by Eq.(\ref{Cxt}). To check the possible size dependence,
we show in Fig.\ref{fig-auto} the time dependence of
$C_\chi (t)$ for the field $H/J=0.5$ on a log-log plot, computed
for (a) $L=16$,  and for (b) $L=20$.
As shown in the figures,
$C_\chi (t)$ shows either a downward curvature
characteristic of the disordered phase, or an upward curvature
characteristic of the long-range ordered phase,
depending on whether the temperature is higher or lower than
$T/J\simeq 0.23$.
Just at $T/J\simeq 0.23$, the linear behavior corresponding to
the power-law decay
is observed. Hence, our  data indicates that
the chiral-glass
transition takes place at $T_{{\rm CG}}/J=0.23(2)$,
in agreement with our above estimate  based on $g_\chi$.
 From the slope of the data at $T=T_{{\rm CG}}$,
the exponent $\lambda $ characterizing the power-law decay of
$C_\chi (t)\approx t^{-\lambda }$ is estimated to be
$\lambda =0.13(2)$. We note that both
our data of $L=16$ shown in Fig.\ref{fig-auto}(a)
and of $L=20$ shown in Fig.\ref{fig-auto}(b)
give almost the same estimates of $T_{{\rm CG}}$ and of $\lambda $, even though
the $L=16$ and $L=20$ data themselves
do not completely overlap,
particularly below $T/J\sim 0.18$.
Anyway,
our observation that $C_\chi (t)$ exhibits an upward curvature below
$T_{{\rm CG}}$, tending
to a nonzero value corresponding to the static
chiral EA parameter $q_{\rm CG}^{\rm EA}$, 
indicates that the chiral-glass ordered state
is ``rigid'' with a nonzero long-range order.
The same analysis  applied to the $H/J=0.1$ case yields
$T_{\rm CG}/J=0.21(2)$ and $\lambda=0.17(2)$.

\begin{figure}
\begin{center}
\includegraphics{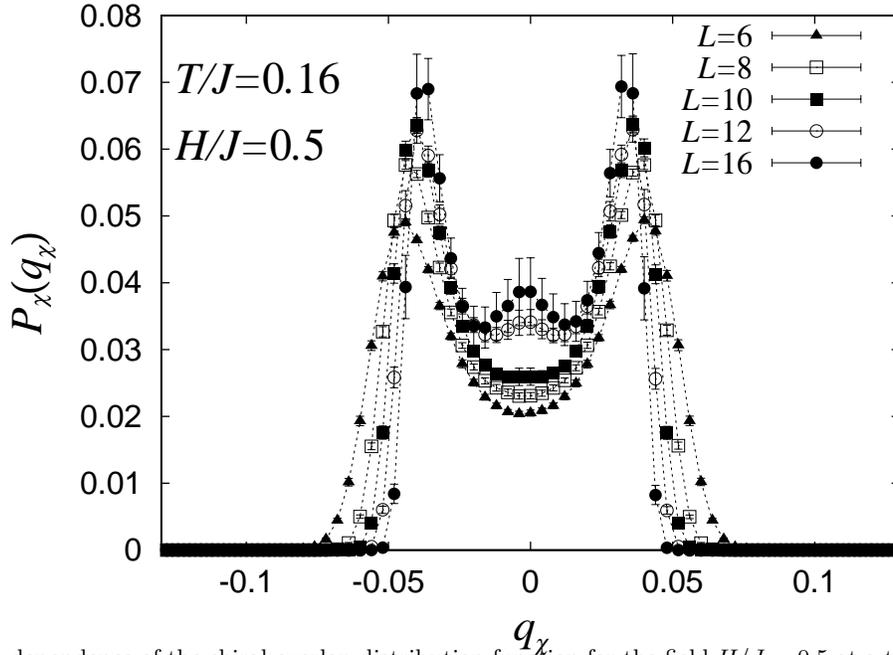}
\caption{The size dependence of the chiral-overlap distribution function
for the field $H/J=0.5$ at a temperature $T/J=0.16$, well below
the chiral-glass transition temperature, $T_{\rm CG}/J
\simeq 0.23$.}
\label{fig-pqxT016}
\end{center}
\end{figure}

In Fig.\ref{fig-pqxT016}, we show the chiral-overlap
distribution function
$P_\chi(q_{\chi})$ for the field $H/J=0.5$
at a temperature $T/J=0.16$, well below $T_{\rm CG}$.
In addition to the standard ``side-peaks''
corresponding to the EA order parameter
$\pm q_{{\rm CG}}^{{\rm EA}}$,
which grow and sharpen with increasing $L$,
there appears a
``central peak'' at $q_\chi =0$ for larger $L$,
which also grows and sharpens with increasing $L$.
The shape of the calculated $P_\chi(q_\chi )$ is
very much similar to the one obtained in Ref.\cite{HK1}
in zero field,
but is quite
different from those
observed in the standard Ising-like models such as the
3D EA model\cite{3DISG} or the mean-field
SK model\cite{SK}. As argued in Ref.\cite{HK1} in case of zero field,
such peculiar  features of $P_\chi(q_\chi )$
are likely to be related to
the {\it one-step\/}-like RSB.
The existence of a negative
dip in the Binder ratio $g_{\chi}$  and the absence of the
standard type of
crossing of  $g_\chi$ at  $g_\chi>0$ are also
consistent with the occurrence of such a one-step-like RSB \cite{HK1,HK2}.
We note that  our data of $P_\chi (q_\chi )$ are
also compatible with
the existence of a continuous plateau between
[$-q_{{\rm CG}}^{{\rm EA}},q_{{\rm CG}}^{{\rm EA}}$]
in addition to the delta-function peaks.

\begin{figure}
\begin{center}
\includegraphics[width=\linewidth]{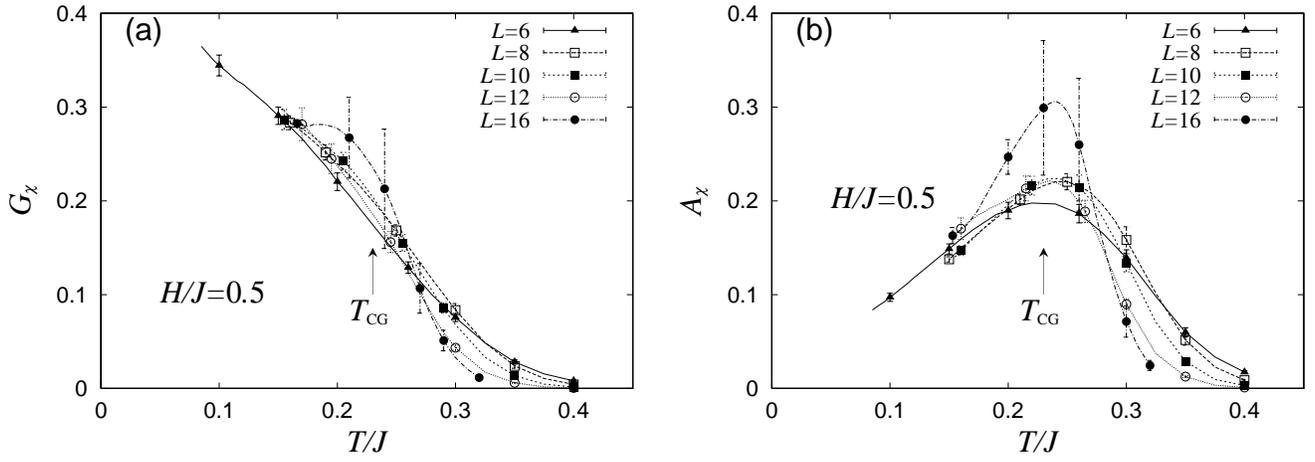}
\caption{Temperature and size dependence of the $G$ and $A$ parameters
of the chirality for the field $H/J=0.5$.}
\label{fig-GAx}
\end{center}
\end{figure}

In Fig.\ref{fig-GAx}, we show the the temperature and size dependence of
the $G_{\chi}$ and $A_{\chi}$
parameters for the field $H/J=0.5$.
Although error bars of
the data are rather large here, the crossing occurs at temperatures
somewhat higher than $T_{{\rm CG}}/J\simeq 0.23$ in both figures, 
while the crossing
temperatures of neighboring sizes ({\it e.g.\/} $L=6$ and $L=8$ {\it etc\/})
gradually shift
towards $T_{{\rm CG}}/J\simeq 0.23$ for larger $L$.
The data are consistent with
our estimate of $T_{{\rm CG}}$ above based on
the Binder ratio and the autocorrelation.
%
%

\newpage

\subsection{Spin-related quantities}
\label{subsecSpin}

\begin{figure}
\begin{center}
\includegraphics[width=\linewidth]{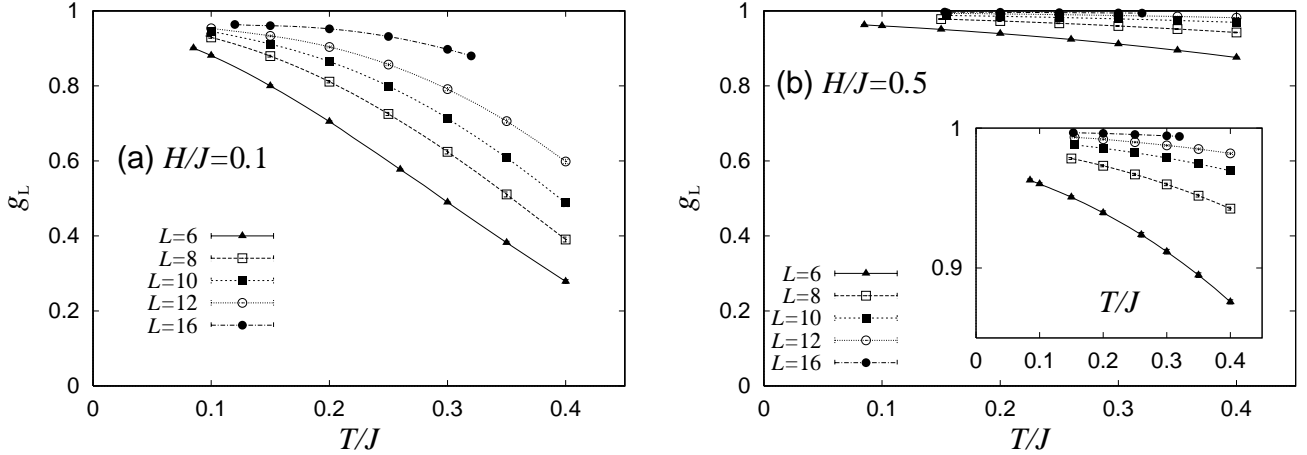}
\caption{Temperature and size dependence of the Binder ratio of the
longitudinal-component of the spin for the fields (a)
$H/J=0.1$, and (b) $H/J=0.5$.
For the field $H/J=0.5$, magnified figure is shown in the inset.}
\label{fig-gsl}
\end{center}
\end{figure}

\begin{figure}
\begin{center}
\includegraphics[width=\linewidth]{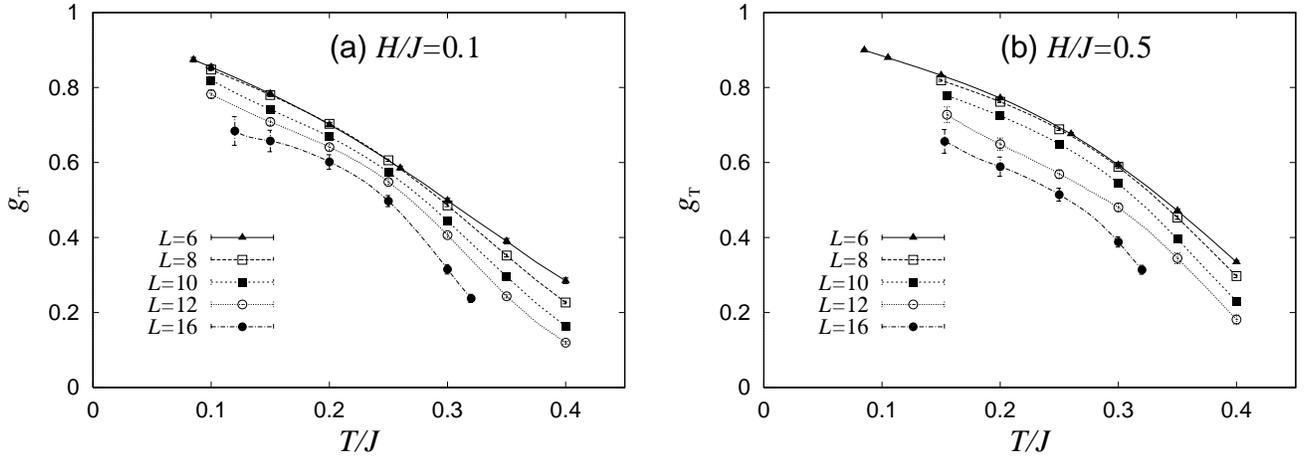}
\caption{Temperature and size dependence of the Binder ratio of the
transverse component of the spin for the fields (a)
$H/J=0.1$,  and (b) $H/J=0.5$.}
\label{fig-gst}
\end{center}
\end{figure}

\begin{figure}
\begin{center}
\includegraphics[width=\linewidth]{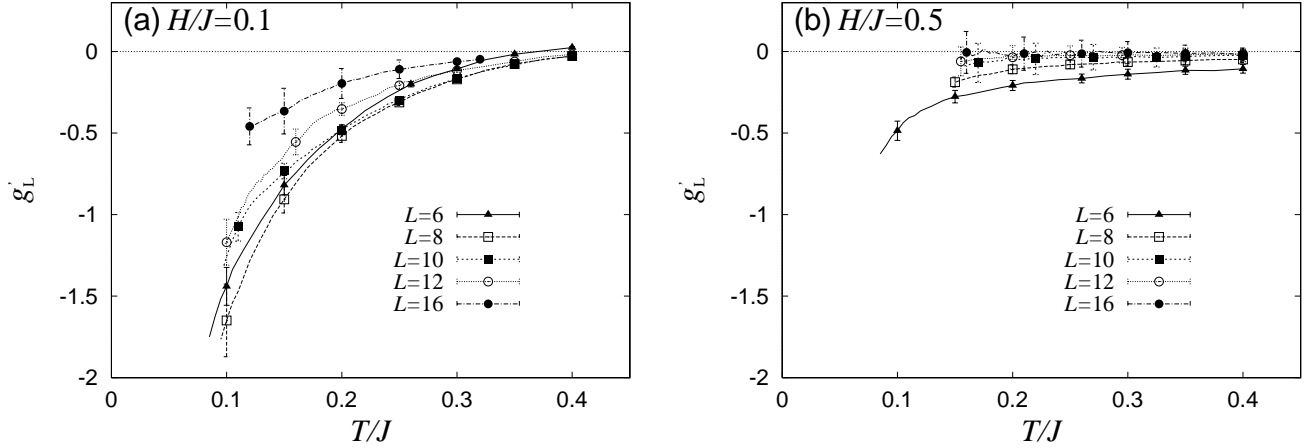}
\caption{Temperature and size dependence of
the connected Binder ratio of the longitudinal component of the spin
for the fields (a) $H/J=0.1$, and (b) $H/J=0.5$.}
\label{fig-gs2}
\end{center}
\end{figure}

In this subsection, we present our MC results of
the spin-related quantities.
In Figs.\ref{fig-gsl} and \ref{fig-gst}, we show
the spin Binder ratios for the
longitudinal and transverse components, respectively, for
the fields (a) $H/J=0.1$, and (b) $H/J=0.5$. For both fields,
the {\it longitudinal\/} Binder ratio $g_{\rm L}$ increases monotonically
toward unity with increasing $L$ at all temperatures studied:
See Fig.\ref{fig-gsl}.
This observation reflects the fact that
the longitudinal component of the spin exhibits
a net magnetization induced by applied fields
at any finite temperatures. By contrast,
the Binder ratio of the
transverse component of the spin
$g_{\rm T}$ decreases toward zero with increasing $L$,
without a negative dip nor a crossing: See Fig.\ref{fig-gst}. This
suggests that the transverse component of spin remains
disordered even below $T_{{\rm CG}}$.
In Fig.\ref{fig-gs2}, we show the
connected Binder ratio of
the longitudinal spin component for the fields (a) $H/J=0.1$, and
(b) $H/J=0.5$. Again,
any anomalous behavior is not appreciable,
no crossing nor extremum. Instead, $g'_{\rm L}$
monotonously approaches zero with increasing $L$, staying negative
at any temperature.
(Strictly speaking,
the data of $L=6$ and $L=8$ for $H/J=0.1$,
exhibits a crossing-like behavior around $T/J\sim 0.2$, but
this is limited to these smaller lattices.)

\begin{figure}
\begin{center}
\includegraphics{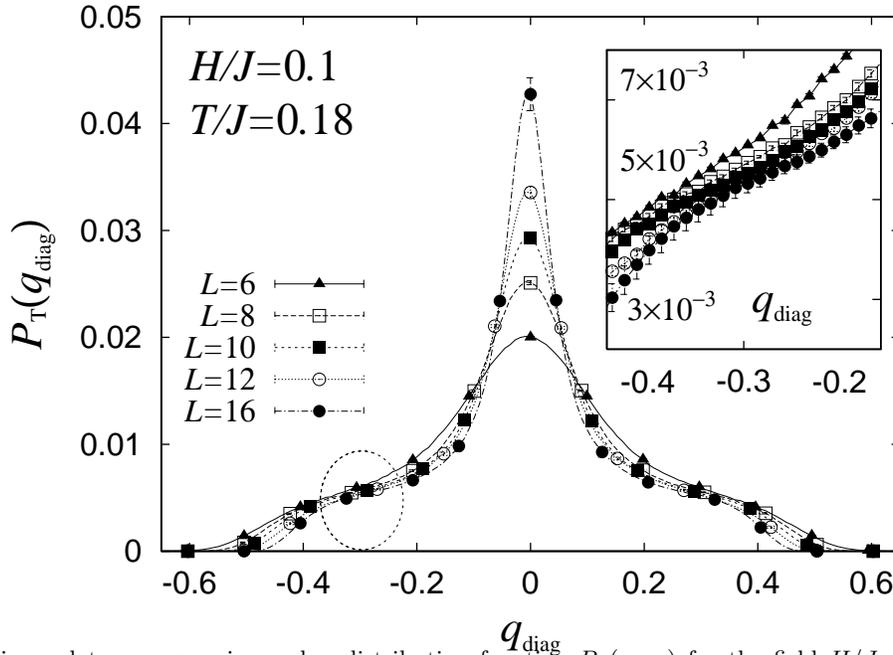}
\caption{The diagonal transverse-spin-overlap distribution function
$P_{\rm T}(q_{{\rm diag}})$
for the field
$H/J=0.1$ at a temperature $T/J=0.18$, well below the chiral-glass
transition temperature $T_{{\rm CG}}/J\simeq 0.21$.
A magnified view of the shoulder part, indicated by the dashed circle
in the main panel,
is shown in the inset.}
\label{fig-qsxy}
\end{center}
\end{figure}

In Fig.\ref{fig-qsxy}, we show
the diagonal transverse-spin-overlap distribution function
$P_{\rm T}(q_{\rm diag})$   for the field
$H/J=0.1$ at a temperature
$T/J=0.18$, well below the chiral-glass transition temperature
$T_{{\rm CG}}=0.21(2)$.
The calculated $P_{\rm T}(q_{\rm diag})$ exhibits a symmetric
``shoulder'' at some nonzero value of $q_{{\rm diag}}$, but as shown
in the inset, this ``shoulder'' gets suppressed with increasing $L$,
{\it not showing a divergent behavior\/}.
Such suppression of the shoulder
indicates that
the chiral-glass ordered state
does not accompany the standard transverse
SG order, at least up to
temperatures $\approx (2/3)T_{{\rm CG}}$. For $H/J=0.5$,
we have also observed similar suppression of the shoulder
up to temperatures as low as around $\approx (2/3)T_{{\rm CG}}$.
Hence, we conclude that the chiral-glass ordered state
does not accompany the standard transverse SG order, at least
just below the chiral-glass transition point.
Strictly speaking, the observed suppression of
the shoulder is  still not inconsistent with
the Kosterlitz-Thouless(KT)-like
critical SG ordered state. However, we note that such a critical SG
ordered state
appearing at $T\leq T_{{\rm CG}}$ is not supported by
our data of $g_{\rm T}$ shown in Fig.\ref{fig-gst}.

\subsection{Critical properties of the chiral-glass transition}
\label{subsecCritical}

   In this subsection, we determine static and dynamical
critical exponents
associated with the chiral-glass transition.
The analysis here is made for the two particular field values,
$H/J=0.1$ and $H/J=0.5$, where most extensive simulations have been
performed. In the analysis below,
we fix $T_{\rm CG}$ to be
$T_{\rm CG}/J=0.21$ ($H/J=0.1$) and
$T_{\rm CG}/J=0.23$ ($H/J=0.5$), as determined above.

\begin{figure}
\begin{center}
\includegraphics[width=\linewidth]{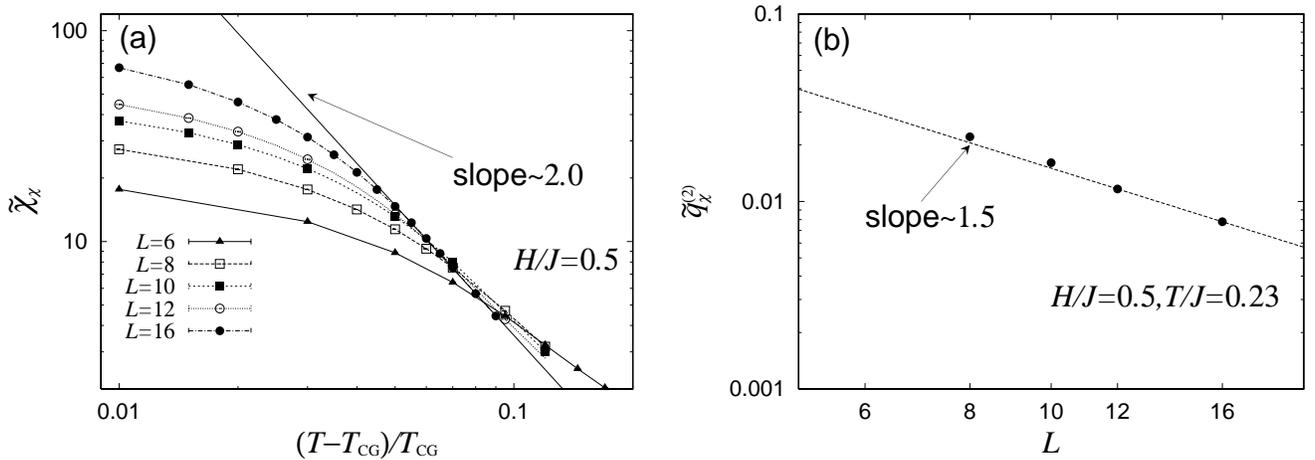}
\caption{(a) Log-log plot of reduced chiral-glass
susceptibility $\tilde{\chi}_\chi$  versus
the reduced temperature for the field $H/J=0.5$.
Its slope $\approx 2.0$ determines
the chiral-glass susceptibility exponent  $\gamma_{\rm CG}=2.0(2)$.
The transition temperature
is assumed here to be $T_{\rm CG}/J=0.23$.
(b) Log-log plot of ${\tilde q}_{\chi}^{(2)}$
versus $L$ 
for the field $H/J=0.5$ at $T/J=0.23\approx T_{\rm CG}/J$.
Its slope $\approx 1.5$ determines the chiral-glass
critical-point-decay exponent to be $\eta_{\rm CG}=0.5(3)$.}
\label{fig-getexponents}
\end{center}
\end{figure}

   We estimate first the chiral-glass susceptibility
exponent $\gamma_{{\rm CG}}$ from the asymptotic slope of the
log-log plot of the reduced
chiral-glass susceptibility $\tilde \chi_\chi$ versus
the reduced temperature $t=(T-T_{\rm CG})/T_{\rm CG}$.
An example is given
in Fig.\ref{fig-getexponents}(a) for the case of
$H/J=0.5$, where an asymptotic slope
$\gamma_{{\rm CG}}=2.0(2)$ is obtained.

We then estimate the chiral-glass
critical-point-decay exponent $\eta_{{\rm CG}}$ from the
$L$-dependence of the chiral-glass order parameter
${\tilde q}_{\chi}^{(2)}$
at $T_{{\rm CG}}$,  according to the relation
${\tilde q}_{\chi}^{(2)}\approx L^{-(1+\eta_{{\rm CG}})}$.
An example  for the $H/J=0.5$ case is shown
in Fig.\ref{fig-getexponents}(b), where we plot
${\tilde q}_{\chi}^{(2)}$ at $T/J=0.23\simeq T_{{\rm CG}}/J$
versus $L$  on a log-log plot.
As can be seen from the figure,
the data lie on a straight line fairly well.
From its slope $\simeq 1.5$, the exponent $\eta_{{\rm CG}}$ is
estimated to be $\eta_{{\rm CG}}=0.5(3)$.

   The rest of the static exponents,
$\alpha_{{\rm CG}}$, $\beta_{{\rm CG}}$, and $\nu_{{\rm CG}}$,
can be estimated from $\gamma _{{\rm CG}}$
and $\eta _{{\rm CG}}$ by using
the standard scaling and hyperscaling relations as
$\alpha_{{\rm CG}}=-1.9(5)$, $\beta_{{\rm CG}}=0.9(4)$
and $\nu_{{\rm CG}}=1.3(3)$.

    The dynamical exponent $z_{{\rm CG}}$ can be estimated from the exponent
$\lambda$,  via the relation
$\lambda=\beta_{{\rm CG}}/z_{{\rm CG}}\nu_{{\rm CG}}$.
 From our above estimate, $\lambda =0.13(2)$, we  get
$z_{{\rm CG}}=5.3(6)$.

   The same procedure is repeated for the case of $H/J=0.1$.
We then get
$\nu_{{\rm CG}}= 1.3(2)$, $\eta_{{\rm CG}}=
0.6(3)$, $z_{{\rm CG}}=4.9(6)$.
These estimates for $H/J=0.1$ agree within errors with the corresponding
estimates for $H/J=0.5$.
Our estimates of the chiral-glass exponents are summarized
in Table \ref{table-criticalexp}, and are compared with
the corresponding zero-field exponents reported in Ref.\cite{HK1}.
The finite-field exponents turn out to
agree within errors with the corresonding
zero-field exponents, suggesting that
the zero-field and finite-field chiral-glass transitions
lie in a common universality class.
We note that this observation is consistent with the chirality
scenario of Refs.\cite{Kawamura92,KawaIma}.
In Table \ref{table-criticalexp}, we also show
the SG exponents
of the 3D Ising EA model\cite{3DISG,3DISGz} together with
typical experimental values (in zero field)
of real Heisenberg-like SG magnet AgMn\cite{expAgMn}.
The critical properties of the chiral-glass
transition differ clearly from those of the 3D Ising EA SG.
By contrast, the chiral-glass exponents are close to the
experimental exponent values for canonical SG AgMn, giving further
support to the
spin-chirality decoupling-recoupling scenario.

%
%
%
%

\newpage

\begin{table}
\begin{center}
\caption{
List of various critical exponents for the chiral-glass transition
in zero- and in finite fields, compared with the corresponding
spin-glass exponents of the 3D Ising EA model \protect\cite{3DISG,3DISGz}
and of real Heisenberg-like SG magnet AgMn
determined experimentally \protect\cite{expAgMn}.}
\label{table-criticalexp}
\includegraphics[width=\columnwidth]{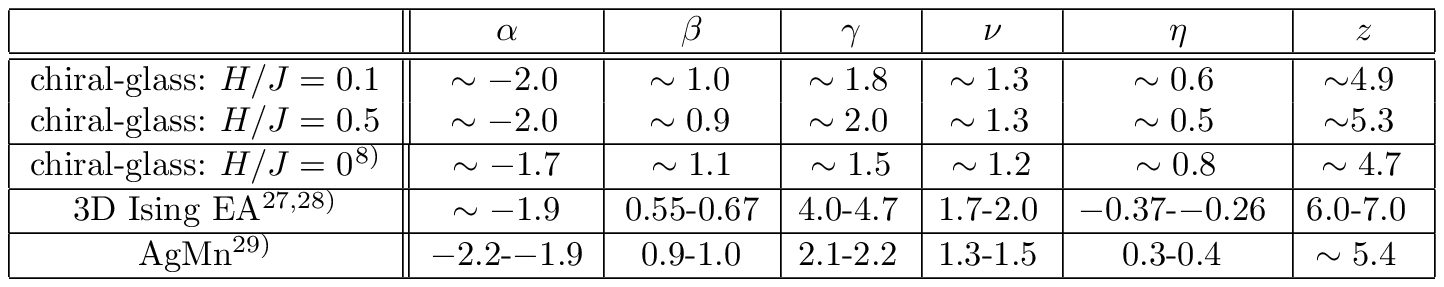}
\end{center}
\end{table}

  As a consistency check of our estimates of exponents and $T_{{\rm CG}}$
values,
we have also done the following:
We use the $\nu _{{\rm CG}}$ value
determined above, $\nu _{{\rm CG}}\sim 1.3$, and
extrapolate the dip temperature of $g_\chi$,
$T_{{\rm dip}}(L)$,
to $L=\infty$ (see the dashed lines  of Fig.\ref{fig-dip}). As mentioned,
such an extrapolation yields the bulk chiral-glass transition
temperature, $T_{{\rm CG}}=0.21(2)$ ($H/J=0.1$) and $T_{{\rm CG}}=0.23(2)$
($H/J=0.5$). These estimates of $T_{{\rm CG}}$
agree with those obtained from  the chiral
autocorrelation and employed in our  scaling analysis.
This guarantees that our analysis of exponents and $T_{{\rm CG}}$
is self consistent.

\begin{figure}
\begin{center}
\includegraphics[width=\linewidth]{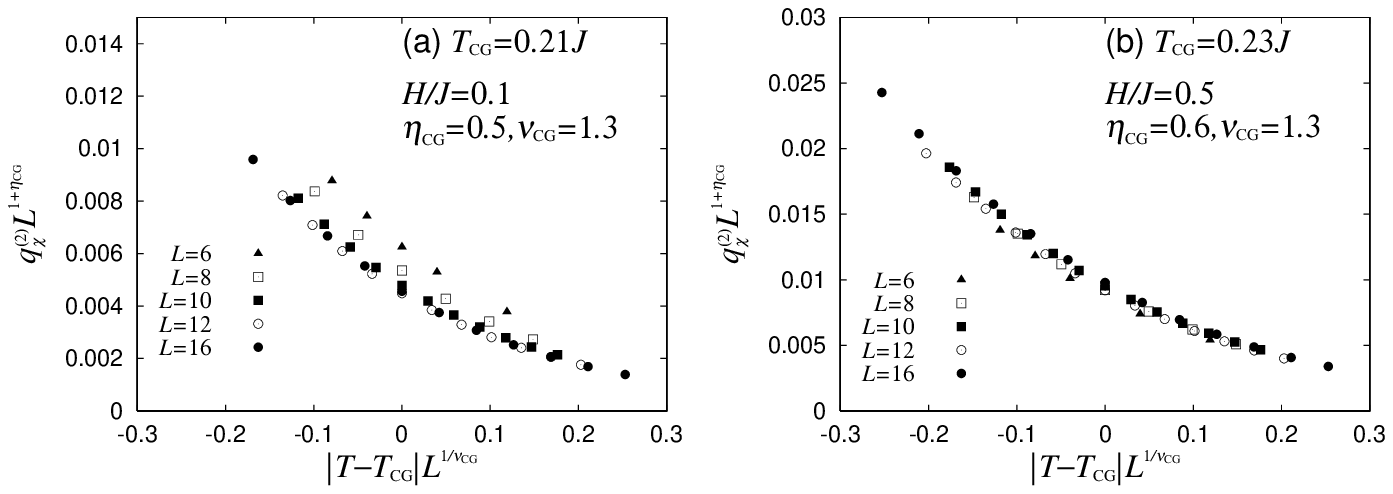}
\caption{Finite-size plot of the chiral-glass
order parameter
for the fields (a) $H/J=0.1$, and (b) $H/J=0.5$.}
\label{fig-scalingC}
\end{center}
\end{figure}

In Fig.\ref{fig-scalingC}, we show the
the standard finite-size
scaling plot for the chiral-glass order parameter
$q_\chi^{(2)}$ based on the relation, 
\begin{equation}
q_\chi^{(2)}\approx L^{-(1+\eta_{{\rm CG}})}f(Lt^{-1/\nu_{{\rm CG}}})\ \ ,
\end{equation}
where the $T_{{\rm CG}}$, $\eta_{{\rm CG}}$ and $\nu_{{\rm CG}}$
values are set to the best values determined above.
As can be seen from the figures,
reasonable data collapsing are obtained, at least for larger lattices.
At the same time, however, one sees that
there exists a systematic deviation from the scaling
for smaller lattices, particularly in the case of $H/J=0.1$. Such a deviation
observed for smaller lattices  suggests the existence of
a significant finite-size correction.

The existence of such significant finite-size effects has
also been suggested
from the behavior of the chiral-overlap distribution function
$P_\chi(q_\chi)$
and of the chiral Binder ratio $g_\chi$.
In a truly asymptotic  critical regime,
$P_\chi(q_{\chi})$ itself should scale at $T=T_{{\rm CG}}$
with tuning one exponent $\eta_{{\rm CG}}$.
However, in the range of sizes studied here $L\leq 16$,
we cannot observe such  a full scaling of $P_\chi(q_{\chi})$.
Such a lack of complete scaling of $P_\chi(q_{\chi})$ gives rise
to certain degrees of uncertainty in our estimate of
$\eta_{{\rm CG}}$: Namely, if one tries to scale the width of the
distribution such as its second moment $q_\chi^{(2)}$,
it yields $\eta_{{\rm CG}}\sim 0.5$ as given above (see Fig.
\ref{fig-getexponents}(b)),
while if one
tries to scale the height of $P_\chi(q_\chi)$, it instead yields
$\eta_{{\rm CG}} \sim 0.4$, which is somewhat smaller than
the above estimate, though still lying within the quoted error bar.
Lack of a complete scaling in $P_\chi(q_{\chi})$ is also reflected
in the behavior of $g_{\chi}$,
which does not show a unique crossing
at $T=T_{\rm CG}$ within the range of sizes studied:
Instead, as shown in Fig.\ref{fig-gx}, the crossing occurs
on the negative side of $g_\chi$ considerably above $T=T_{{\rm CG}}$,
while the crossing points tend to come down toward
$T_{{\rm CG}}$ as $L$ increases.

\begin{figure}
\begin{center}
\includegraphics[width=\linewidth]{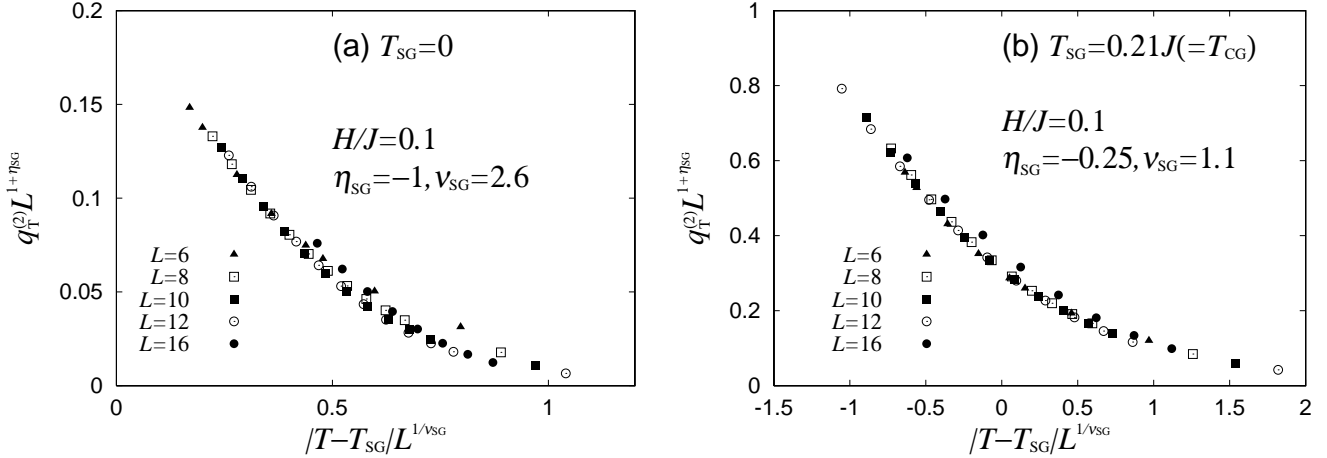}
\caption{Finite-size scaling plot
of the transverse spin-glass order parameter for the field $H/J=0.1$,
assuming (a) $T_{\rm SG}=0$, and
(b) $T_{{\rm SG}}=0.21J(=T_{{\rm CG}})$.
}
\label{scaleSph01}
\end{center}
\end{figure}

\begin{figure}
\begin{center}
\includegraphics[width=\linewidth]{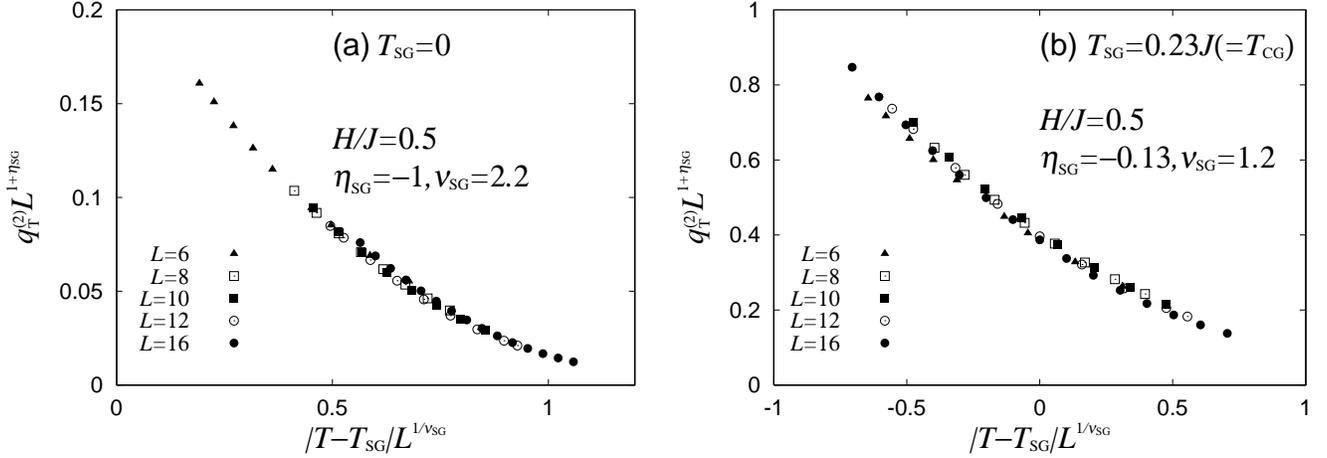}
\caption{Finite-size scaling plot
of the transverse spin-glass order parameter for the field $H/J=0.5$,
assuming (a) $T_{\rm SG}=0$, and
(b) $T_{{\rm SG}}=0.23J(=T_{{\rm CG}})$.
}
\label{scaleSph05}
\end{center}
\end{figure}

Concerning the transverse spin order, from the behaviors of
the Binder ratio and of the diagonal transverse-spin-overlap
distribution function, we have already found a strong numerical evidence that
the chiral-glass transition does not
accompany the transverse SG order, at least just below
$T_{{\rm CG}}$.
In other words, the transverse component of the spin orders only
at zero temperature, or else, if it orders at a finite temperature,
the associated transverse SG transition temperature $T_{{\rm SG}}$
is significantly lower than $T_{{\rm CG}}$, say,
below $\approx (2/3)T_{{\rm CG}}$. We warn the reader here  that,
so long as one looks at the SG correlation or the
SG order parameter, a rather
careful analysis is required to really see such a behavior.
As an example,
we show in Fig.\ref{scaleSph01} the standard finite-size scaling plots of the transverse
SG order parameter $q^{(2)}_{\rm T}$ for $H/J=0.1$; (a)
the one assuming $T_{{\rm SG}}=0$,
and (b) the other assuming $T_{{\rm SG}}=0.21J(=T_{{\rm CG}})$.
Similar plots are given in Fig.\ref{scaleSph05}
for the field $H/J=0.5$ with assuming $T_{{\rm SG}}=0$, assuming
(a) $T_{{\rm SG}}=0$ and (b) $T_{{\rm SG}}=0.23J(=T_{{\rm CG}})$.
At a look, both fits seem equally acceptable without
appreciable difference if the exponents are adjusted in appropriate ways.
Then, one may wonder if
the transverse spin might order
simultaneously with the chirality, with the associated
SG exponents
$\nu _{{\rm SG}}\simeq 1.1(\approx \nu _{{\rm CG}}$) and
$\eta _{{\rm SG}}\simeq -0.25$: See Fig.\ref{scaleSph01}.
We believe, however,
this not to be the case due to the following reasons.
First,  as shown above,
such simultaneous chirality and
transverse-spin ordering contradicts with our result
of the Binder ratio $g_{\rm T}$
and the diagonal transverse-spin-overlap distribution
function $P_{\rm T}(q_{{\rm diag}})$.
Second, a closer inspection of the data
reveals that, at and below $T=T_{{\rm CG}}$,
there exists an
important difference between the behaviors of the transverse spin
$q_{\rm T}^{(2)}$
and of the chirality $q_{\chi}^{(2)}$.

\begin{figure}
\begin{center}
\includegraphics[width=\linewidth]{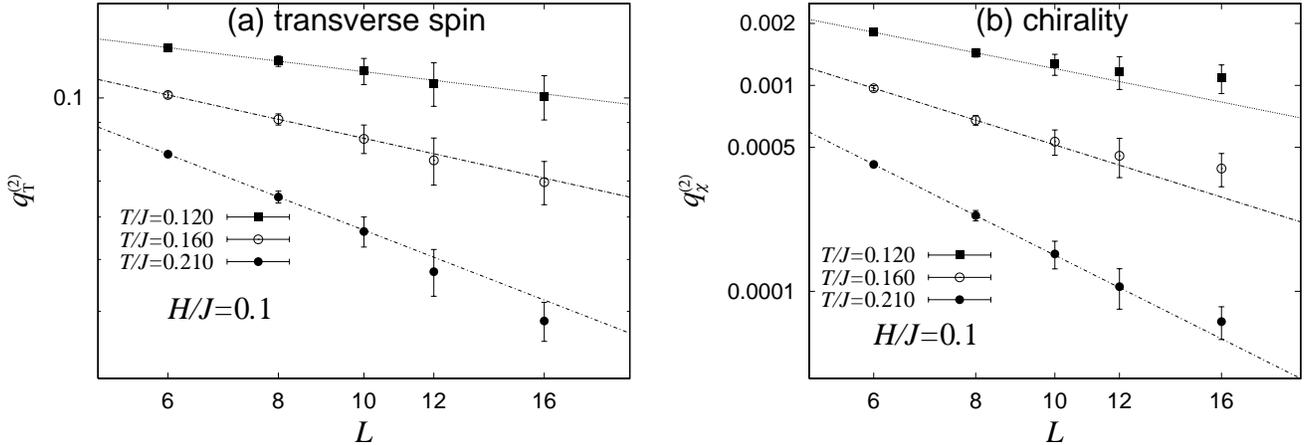}
\caption{Size dependence of the  order parameter
for the field $H/J=0.1$,
at several temperatures at and below
$T_{\rm CG}$: (a) the transverse-spin-glass order parameter
$q_{\rm T}^{(2)}$ , and (b) the chiral-glass order parameter
$q_\chi^{(2)}$. The chiral-glass transition temperature at this
field is $T_{{\rm CG}}/J\simeq 0.21$.
To emphasize the deviation from the linearity,
lines connecting the two small-size data
$L=6$ and $8$ are drawn at each temperature.
}
\label{fig-q2log}
\end{center}
\end{figure}

In Fig.\ref{fig-q2log}, we show on a
log-log plot the size dependence of
the both order parameters, $q_{\rm T}^{(2)}$ and $q_{\chi}^{(2)}$,
at several temperatures at and below $T_{{\rm CG}}$ . As can be seen
from Fig.\ref{fig-q2log}(a),
below $T_{{\rm CG}}$ the chiral-glass order parameter
exhibits a clear upbending
for larger $L$,
indicating that $q_{\chi}^{(2)}$ tends to a nonzero value
in the thermodynamic limit. In sharp contrast to this,
such an upbending is never seen in the transverse SG order
parameter $q_{\rm T}^{(2)}$:
Instead,
$q_{\rm T}^{(2)}$ shows a slight downbedning
behavior at $T=T_{{\rm CG}}$,
which gradually shifts to the near linear
behavior at lower temperatures. The observed behavior of $q_{\rm T}^{(2)}$
is consistent
with either, (a) the onset of the Kosterlitz-Thouless(KT)-like
transition
at a finite temperature below which the spin-glass correlations
decay algebraically with a power-law or, (b) the
gradual growth of the transverse SG correlation length $\xi $
which exceeds the
investigated system size $L=16$
around a certain nonzero temperature close to $T_{{\rm CG}}$.
In the former case,
there should exist a well-defined finite SG transition temperature
with the critical SG ordered state,
while, in the latter case, there need not be a thermodynamic
SG transition at a finite temperature.
Generally speaking, it is difficult to discriminate between
the above two possibilities only from the $q_{\rm T}^{(2)}$ data
of finite sizes with $L\leq \xi$.

Nevertheless, we believe  we can at least exclude here the possibility
that the KT-like transverse SG transition occurs {\it simultaneously\/}
with the chiral-glass transition  at $T=T_{{\rm CG}}$, accompanied by
the critical SG ordered state at $T\leq T_{{\rm CG}}$.
First, we note that such a critical SG ordered state
is not supported by
our data of $g_{\rm T}$ of Fig.\ref{fig-gst}. Second,
the transverse {\it spin}-glass correlation-length exponent estimated
in Figs.\ref{scaleSph01}(b) and \ref{scaleSph05}(b)
assuming the simultaneous spin and chiral
transition, $\nu\simeq 1.1$, is far from from the
lower-critical-dimension (LCD) value, $\nu=\infty$, generically
expected for such a KT-like transition. In so far as one
insists that the transverse SG order occurs simultaneously with
the chiral-glass order, our numerical estimate of the
transverse SG correlation-length exponent
is not compatible with the LCD value $\nu =\infty$,
which is now hard to reconcile with
the KT-like behavior observed in $q_{\rm T}^{(2)}$ at $T\leq
T_{{\rm CG}}$.

In fact, as recently argued in Ref.\cite{KawaLi} for the case of the 
{\it XY\/}SG, the chirality scenario gives
the possible cause why
simultaneous spin and chiral orderings are apparently observed
in  the SG order parameter or
the SG correlation function.
This would closely be related to the
length and time scales of the measurements.
Here, one should be aware of the fact
that the spin-chirality decoupling is a {\it long-scale\/} phenomenon:
At short scale, the chirality is never independent of the spin by its
definition, roughly being its squared ($\chi \sim S_{\rm T}^2S_{\rm L}
\sim S_{\rm T}^2$)
as expected from the naive power counting.
Hence, the behavior of the spin-correlation related quantities,
including the SG
order parameter which is a summed correlation,  might well reflect
the critical singularity associated with the {\it chirality\/}
{\it i.e.\/}, the one of the chiral-glass transition,
up to certain length and time scale.
In such a scenario, apparent (not true) transverse ``spin-glass
exponents''
expected would be $\nu '_{{\rm SG}}\sim \nu_{{\rm CG}}\sim
1.3$ and $\eta '_{{\rm SG}}\sim -0.25$,
the latter being derived from the short-scale
relation, $1+\eta_{{\rm CG}}\sim 2(1+\eta'_{{\rm SG}})$.
Note that these values are not very far from the ones we get from
the finite-size scaling analysis of
Figs.\ref{scaleSph01}(b) and \ref{scaleSph05}(b), assuming the
simultaneous occurrence of the spin and chiral transition.
However, we stress again that such a disguised
criticality in the spin sector is only  a short-scale phenomenon, not a true
critical one.

%
%

%
%

%
%
%

%
\subsection{Phase diagram}
\label{subsecPhase}

\begin{figure}
\begin{center}
\includegraphics[width=\linewidth]{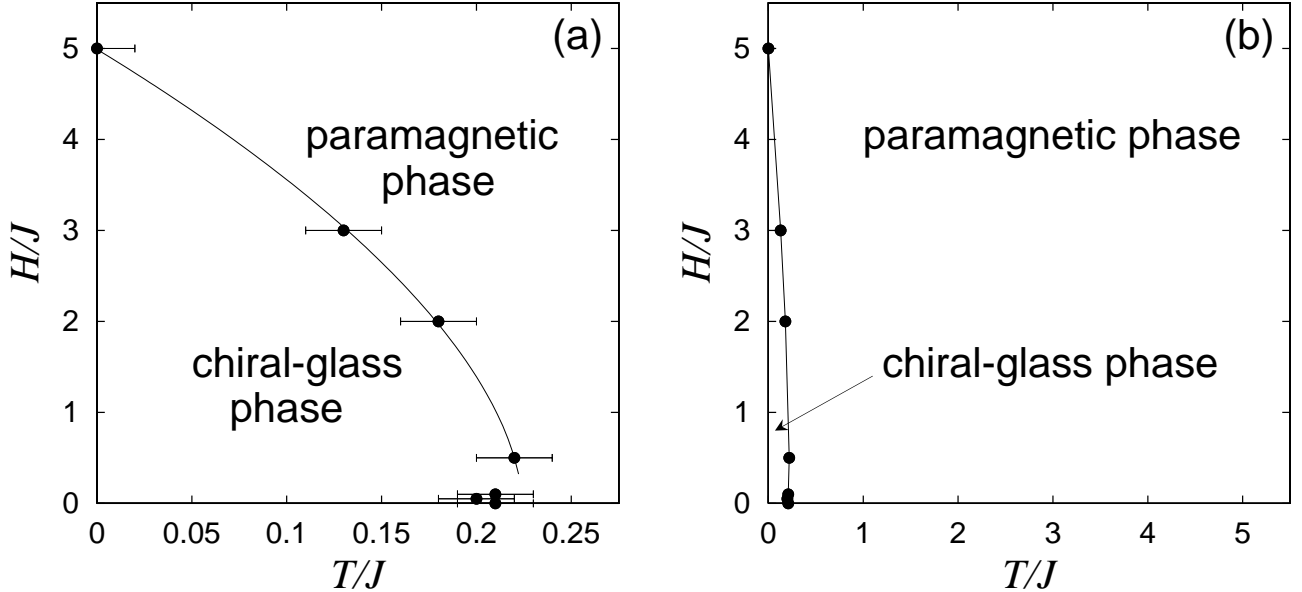}
\caption{$H$-$T$ phase diagram of the 3D $\pm J$ Heisenberg SG determined
by the present simulation.
Note that the energy scales of the $H$ and of the $T$ axes are
mutually different
in Fig.(a),
while they are taken to be common in Fig.(b).}
\label{fig-phase}
\end{center}
\end{figure}

By collecting our estimates of the $T_{{\rm CG}}$ values
for various field values, as obtained by the extrapolation of
$T_{{\rm dip}}(L)$ to $L=\infty$,
we construct a phase
diagram in the temperature vs. magnetic field plane. The result is shown
in Fig.\ref{fig-phase}.
We have used here the zero-field estimate of Ref.\cite{HK3},
$T_{{\rm CG}}/J=0.21(2)$.
Error bars are estimated here from the differences between
the extrapolated $T_{{\rm CG}}$ values via
the $1/L$ and $1/L^{1/1.3}$ fits.
As is evident from Fig.\ref{fig-phase},
the chiral-glass state
remains quite robust against magnetic fields. This is most evident in
Fig.\ref{fig-phase}(b) where we draw the same phase diagram on a plot where
both the
temperature and the magnetic-field axes have common energy scale.
Indeed,
$T_{{\rm CG}}(H)$ is not much reduced from the zero-field value
even at a field as large as ten times of $T_{{\rm CG}}(0)$.
At lower fields, the chiral-glass
transition line is almost orthogonal to the $H=0$ axis,
consistent with the behavior Eq.(\ref{eqn:phaseline})
derived from the chirality scenario.
Our data are
even not inconsistent with the coefficient $c$
in Eq.(\ref{eqn:phaseline})
being slightly
negative so that
$T_{{\rm CG}}(H)$ initially {\it increases\/} slightly with $H$,
though it is difficult to draw a
definite conclusion due to the scatter
of our estimate of $T_{{\rm CG}}(H)$.
If one remembers here our MC observation that the application of
a weak magnetic field tends to increase the mean local
amplitude of the chirality, $\bar \chi$, from its zero-field value,
such an initial increase of $T_{{\rm CG}}(H)$ seems not totally unlikely.

%
%
%
\section{Summary and Discussion}
\label{summary}

In summary, we have performed large-scale  equilibrium Monte Carlo
simulations on the 3D isotropic
Heisenberg SG in finite magnetic fields.
We have confirmed that our MC results are consistent with the
chirality scenario of Ref.\cite{Kawamura92}.
Among other things, we have verified
the occurrence of a finite-temperature chiral-glass
transition in applied fields,
essentially of the same character as the zero-field one.
The chiral-glass ordered state exhibits a one-step-like peculiar
RSB, while it does not accompany the transverse SG order, at least
up to temperatures around $\approx (2/3)T_{{\rm CG}}$. The criticality
of finite-field chiral-glass transitions seems to be common with
that of the  zero-field one, which, however,
clearly differs from the criticality of the
standard 3D Ising EA model. Meanwhile, the chiral-glass exponents
turn out to be close to the experimental
exponents determined for canonical SG such as AgMn.
We have also constructed a magnetic phase diagram of the 3D Heisenberg SG
model.
The chiral-glass transition line in the $H$-$T$ plane is found
to be almost vertical to
the temperature axis, up to rather high fields of order
$H\sim 10T_{\rm CG}(0)$, indicating that the chiral-glass ordered state
is quite robust against magnetic fields.
This somewhat surprising property probably
arises from the fact that
the magnetic field
couples in the Hamiltonian directly fo the spin,
{\it not to the chirality\/},
and the effective coupling between the field and the chirality is rather weak.
The chiral-glass transition line has  a character of the
Gabay-Toulouse line of the mean-field model, yet its physical origin
being entirely different.

\begin{figure}[b]
\begin{center}
\includegraphics{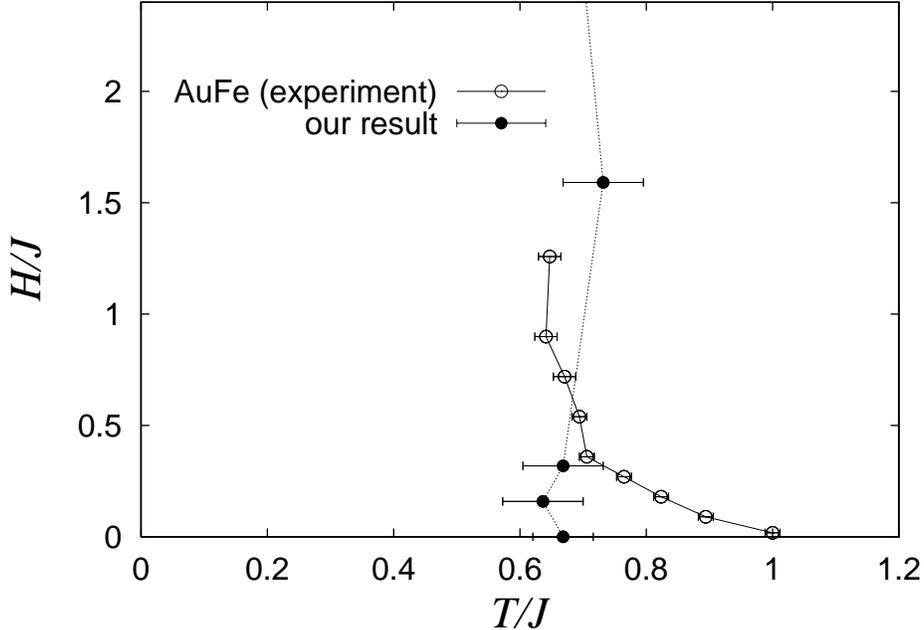}
\caption{
Experimental phase diagram of AuFe
as determined by Campbell {\it et al\/} by means of
torque measurements [Ref.\protect\cite{Campbell}].
For comparison, we also
show the present numerical result of the magnetic
phase diagram of the 3D $\pm J$
Heisenberg SG model.
The way how we scale the units of magnetic field and
temperature in plotting the experimental data is explained in the text.
}
\label{phasedg}
\end{center}
\end{figure}

   It is not immediately possible to make a direct
comparison of our results with experiments. This is mainly because the
random magnetic anisotropy, which inevitably exists in real SG
materials, is not introduced in our present model.
Furthermore, in real SG magnets,  spins do
not necessarily sit
on a simple-cubic lattice, nor interact with other spins
via the nearest-neighbor $\pm J$ coupling, {\it etc\/}.
In spite of these obvious limitations, it might be interesting to try to
compare our present magnetic phase diagram with the experimental one for
Heisenberg-like SG magnets.
Chirality scenario claims that, in the high-field region where the
anisotropy is negligible relative to the applied magnetic field,
the SG transition line should essentially
be given by the chiral-glass transition
line of the fully isotropic system.
If so, our present result entails that the SG
transition line of real Heisenberg-like SG should be
almost vertical against the temperature axis in the high-field
regime where  the magnetic field overwhelms the random magnetic anisotropy.
In Fig.\ref{phasedg}(a), we reproduce the experimental $H$-$T$ phase diagram of
canonical SG AuFe from Ref.\cite{Campbell}.
In the same
figure, we also show our  present result of
the chiral-glass transition line, scaled in the following way.
We try to mimic the real system by the classical Heisenberg Hamiltonian
with an effective coupling $J$ and an effective magnetic field $H$,
which is defined in terms of Eq.(\ref{eqn:phaseline}).
First, we estimate the 
zero-field transition temperature of the  hypothetical 
{\it isotropic\/} system to be
$T_g\approx 10$K, by extrapolating the
high-field GT-like transition line of AuFe to  $H=0$. Then, with the
knowledge of our present estimate of $T_{{\rm CG}}\approx 0.2J$,  
we estimate the relevant $J$ roughly to be  50K.
The field intensity $H$ is then
translated into the field intensity in the
standard unit $H^*$
by the relation $H=p_{\rm eff}H^*$,
$p_{\rm eff}$ being the effective  Bohr number:
In case of AuFe, $p_{\rm eff}$ was experimentally estimated
to be $4.55\mu _B$, where $\mu _B$ is the Bohr magneton\protect\cite{CWlaw}.
Thus, our Fig.\ref{fig-phase}
suggests that the SG phase boundary of AuFe might stay
nearly vertical
up to the field as high as
$H\sim 10T_{\rm SG}(0)\sim 40$[T]. Of course,
considering the difference in microscopic
details between the present model and real AuFe, one cannot
expect a truly quantitative correspondence here.
Anyway, further high-field experiments on AuFe and other Heisenberg-like
SG magnets might be worthwhile to determine the SG phase boundary
in the high-field regime.

   In order to make further comparison with the experimental phase diagram
in the low-field regime,
it is essential to examine the effects of random magnetic anisotropy
inherent to real SG materials. Indeed, in the low-field regime where
the applied field intensity is comparable to or weaker than the random
magnetic anisotropy, the chirality scenario predicts the appearance of
a singular crossover line which has some character of the AT-line of the
mean-field model\cite{KawaIma,Kawaunpub}.
In order to make further insight into
the spin-glass and the chiral-glass
orderings in magnetic fields and
to check further the validity of the chirality scenario,
it would be interesting to make similar finite-field simulations
for the {\it anisotropic\/}
3D Heisenberg SG model.

\section*{Acknowledgements}

    The numerial calculation was performed on the HITACHI SR8000
at the supercomputer system, ISSP, University of Tokyo.
The authors are
thankful to Dr. K. Hukushima, Dr. H. Yoshino and 
Dr. I. A. Campbell for useful discussion.

\end{document}